\documentclass{IEEEtaes}

\usepackage{color,array,amsthm}
\usepackage{graphicx}

\jvol{XX}
\jnum{XX}

\usepackage{array}
\usepackage{amsthm,amsmath,amssymb}
\usepackage{mathrsfs}
\usepackage{multirow}
\usepackage{multicol}
\usepackage{siunitx}

\usepackage{booktabs}
\usepackage{graphicx}
\usepackage{enumerate}
\usepackage{subcaption}
\usepackage{cite}
\usepackage{color}
\theoremstyle{plain}
\newtheorem{thm}{\protect\theoremname}
\theoremstyle{plain}
\newtheorem{lem}[thm]{\protect\lemmaname}
\usepackage{babel}
\providecommand{\lemmaname}{Lemma}
\providecommand{\theoremname}{Theorem}
\begin{document}
\newcommand{\ee}{\end{equation}}
\newcommand{\br}{{\mbox{\boldmath{$r$}}}}
\newcommand{\bp}{{\mbox{\boldmath{$p$}}}}
\newcommand{\bpi}{\mbox{\boldmath{ $\pi $}}}
\newcommand{\bn}{{\mbox{\boldmath{$n$}}}}
\newcommand{\balfa}{{\mbox{\boldmath{$\alpha$}}}}
\newcommand{\ba}{\mbox{\boldmath{$a $}}}
\newcommand{\bta}{\mbox{\boldmath{$\beta $}}}
\newcommand{\bg}{\mbox{\boldmath{$g $}}}
\newcommand{\bPsi}{\mbox{\boldmath{$\Psi $}}}
\newcommand{\bsigma}{\mbox{\boldmath{ $\Sigma $}}}
\newcommand{\bGamma}{{\bf \Gamma }}
\newcommand{\bA}{{\bf A }}
\newcommand{\bP}{{\bf P }}
\newcommand{\bX}{{\bf X }}
\newcommand{\bI}{{\bf I }}
\newcommand{\bR}{{\bf R }}
\newcommand{\bZ}{{\bf Z }}
\newcommand{\bz}{{\bf z }}
\newcommand{\bx}{{\mathbf{x}}}
\newcommand{\bM}{{\bf M}}
\newcommand{\bU}{{\bf U}}
\newcommand{\bD}{{\bf D}}
\newcommand{\bJ}{{\bf J}}
\newcommand{\bH}{{\bf H}}
\newcommand{\bK}{{\bf K}}
\newcommand{\bm}{{\bf m}}
\newcommand{\bN}{{\bf N}}
\newcommand{\bC}{{\bf C}}
\newcommand{\bL}{{\bf L}}
\newcommand{\bF}{{\bf F}}
\newcommand{\bv}{{\bf v}}
\newcommand{\bSigma}{{\bf \Sigma}}
\newcommand{\bS}{{\bf S}}
\newcommand{\bs}{{\bf s}}
\newcommand{\bO}{{\bf O}}
\newcommand{\bQ}{{\bf Q}}
\newcommand{\btr}{{\mbox{\boldmath{$tr$}}}}
\newcommand{\bNSCM}{{\bf NSCM}}
\newcommand{\barg}{{\bf arg}}
\newcommand{\bmax}{{\bf max}}
\newcommand{\test}{\mbox{$
	\begin{array}{c}
		\stackrel{ \stackrel{\textstyle H_1}{\textstyle >} } { \stackrel{\textstyle <}{\textstyle H_0} }
	\end{array}
	$}}
\newcommand{\tabincell}[2]{\begin{tabular}{@{}#1@{}}#2\end{tabular}}
\newtheorem{Def}{Definition}
\newtheorem{Pro}{Proposition}
\newtheorem{Lem}{Lemma}
\newtheorem{Exa}{Example}
\newtheorem{Rem}{Remark}
\newtheorem{Cor}{Corollary}
\renewcommand{\labelitemi}{$\bullet$}

\title{The Trajectory PHD Filter for\\ Coexisting Point and Extended Target Tracking}

\author{Shaoxiu~Wei}
\affil{University of Electronic Science and Technology of China, China} 

\author{Ángel F. García-Fernández}
\affil{University of Liverpool, UK} 

\author{Wei Yi}
\affil{University of Electronic Science and Technology of China, China}

%% \author{FOURTH D. AUTHOR}
%% \affil{University of Colorado, Colorado, USA}

\authoraddress{S. X. Wei and W. Yi are with the School of Information and Communication Engineering, University of Electronic Science and Technology of China. (e-mail: sxiu\_wei@hotmail.com; kusso@uestc.edu.cn). Angel F. García-Fernández is with the Department of Electrical Engineering and Electronics, University of Liverpool, Liverpool L69 3GJ, U.K. (e-mail: angel.garcia-fernandez@liverpool.ac.uk).}

%\markboth{SHAOXIU WEI ET AL.}{SHORT ARTICLE TITLE}
\maketitle

\begin{abstract}
This paper develops a general trajectory probability hypothesis density (TPHD) filter, which uses a general density for target-generated measurements and is able to estimate trajectories of coexisting point and extended targets. First, we provide a derivation of this general TPHD filter based on finding the best Poisson posterior approximation by minimizing the Kullback-Leibler divergence, without using probability generating functionals. Second, we adopt an efficient implementation for this filter, where Gaussian densities correspond to point targets and Gamma Gaussian Inverse Wishart densities for extended targets. Simulation and experimental results show that the proposed filter is able to classify targets correctly and obtain accurate trajectory estimation.
\end{abstract}


\section{INTRODUCTION}
{A}{\scshape utonomous} vehicles promise the possibility of fundamentally changing the transportation industry, with an increase in both highway capacity and traffic flow \cite{Auto-Driving2012,Autodriving2015}. It is required to simultaneously extract the environmental information that incorporates dynamic as well as static objects through road infrastructure and other vehicles. Accordingly, the multi-target tracking (MTT) approaches are widely used for estimating the states and number of dynamic targets, which may appear, move and disappear, given noisy sensor measurements in time sequence\cite{Blackman1986book}. %The MTT problem can be solved in a Bayesian framework by computing the posterior density on the current set of targets, given probabilistic models for target births, dynamics and deaths, and also models for the measurements, obtained from one or multiple sensors. %The target birth model contains probabilistic information on where targets may appear in the surveillance area, and it enables the resulting filters to contain information on potential targets that may remain occluded.  
\par There are two main kinds of approaches to solve MTT problems. The first category is based on random vectors, such as the
joint probabilistic data association (JPDA) filter \cite{JPDA-1998,JPDA-1983} and the
multiple hypotheses tracking (MHT) \cite{Blackman1986book,Blackman2004MHT}. % and  graph-based filters \cite{Meyer2020Graph, Meyer2021Graph}.
The second category is based on random finite set (RFS)\cite{Mahler2003RFS,Mahler2007RFSbook,Mahler2014RFSbook,Vo2013GLMB}, which is closely related to stochastic geometry \cite{Haenggi09} and stochastic flows \cite{Bakut76}. Among them, RFS-based algorithms have been proved to possess an excellent tracking performance in various scenarios\cite{TvT-slam,Reuter2014LMB,Angel2018PMBM,TvT-Shaoxiu}. The probability hypothesis density (PHD) filter, known for its low computational burden among all RFS based filters (including labelled RFS approaches \cite{Vo2013GLMB}), possesses a high efficiency in solving real time tracking problem\cite{Mahler2003RFS, Vo2006PHD}. It propagates the first-order multi-target moment, also called intensity, through prediction and update step. The PHD filter can also be derived by propagating a Poisson multi-target density through the filtering recursion, obtained via Kullback-Leibler divergence (KLD) minimization\cite{Mahler2003RFS, Angel2015KLD}.

\par In MTT, an important topic is to obtain accurate trajectory estimates and mitigate trajectory fragmentation. Calculating or approximating the posterior over a set of trajectories \cite{Svensson2014TMTT,Angel2019TPHD,Granstrom2018TPMBM,Angel2020TMB,Angel2020TMTT} is a efficient approach to meet the above requirements. Among these approaches, the trajectory PHD (TPHD) filter \cite{Angel2019TPHD} extends the PHD filter to estimate trajectories from first principles using trajectory
RFSs, and inherits the low computational burden associated to the PHD filter. The TPHD filter propagates a PPP distribution over the set of alive trajectories through the filtering recursion\cite{Angel2015KLD, TRFS1}. To derive TPHD filters, the theory of sets of trajectories \cite{Angel2020TMTT} is needed, e.g., it is required to use the set integral for sets of trajectories, the Bayesian recursion for sets of trajectories, and the concept of PPP over sets of trajectories. The TPHD filter propagates the best Poisson multi-trajectory density under the standard point target dynamic and measurement models\cite{Angel2019TPHD}. The Gaussian mixture is proposed to obtain a closed-form solution of the TPHD filter. Other trajectory-based filters for point targets are the trajectory multi-Bernoulli filter, the trajectory PMBM filter and the trajectory PMB filter \cite{Angel2020TMB,Granstrom2018TPMBM, Angel-2020-TPMB}. 

%Based on sets of trajectories, these filters  contain all information to answer trajectory-related questions, which are of major importance in autonomous vehicles and smart traffic systems. 

\par In order to develop Bayesian filters, we need to model the distribution of the measurements given the targets as well as clutter. There are two main types of modeling for target-generated measurements: the point target model and the extended target model. In general, a point target model, in which each target can at most generate one measurement, is applied for targets that are smaller than the sensor resolution, given its size and distance from the sensor\cite{Angel2018PMBM,Granstrom2018TPMBM}. Conversely, in an extended target model, a target may generate more than one measurement, which can happen if the target occupies multiple resolution cells of the sensor \cite{Extracking-overview2017,Meyer2021Graph,extracking1,Simon2023ex,extracking2,GLMB-extracking1,GLMB-extracking3}. The Poisson point process (PPP) is widely used in the measurement model for an extended target\cite{Extracking-Poissonmeas2005}, i.e., a Poisson distributed random number of measurements are generated, distributed around an extended target at each time step. There are extended target filters for sets of targets \cite{Grans2012exPHD1,Grans2012exPHD2,Grans2012exCPHD,ExPMBM2020,GransXradar2015,Extracking-overview2017} and also sets of trajectories \cite{ExTPMB,ExTPHD}. Except for extracting dynamic target state information, the mean number of generated measurements from an extended target or the size of the target are also modeled in these filters. Besides, there are also approaches for Bayesian smoothing of target extent based on random matrix model\cite{Exsmooth,ExTPHD}.

\par There are many applications in which it is important to develop models and MTT algorithms for scenarios in which there are simultaneous point and extended targets\cite{Mahler2014RFSbook, ExpPMBM2021}. For example, in a self-driving vehicle application, pedestrians may be modeled as point targets while some vehicles are taken as extended targets. Besides, the distinction between point and extended targets may also depend on the distance to the sensor. 
\begin{figure}[!t]
	\centering
	\includegraphics[width=2.8in]{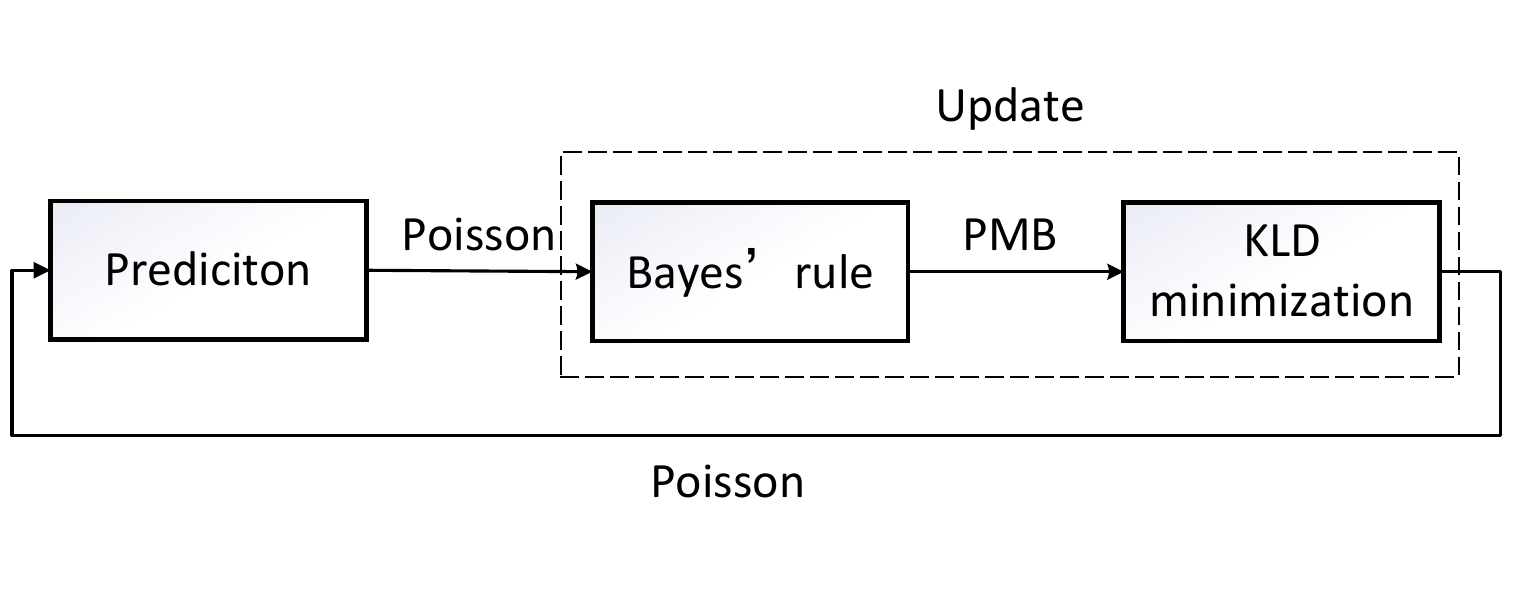}
	\caption{Diagram of the proposed TPHD filter, with general target-generated measurements, Poisson clutter, Poisson birth model and no spawning. This filter propagates a Poisson density on the set of alive trajectories, whose form is kept in the prediction. Bayes' rule provides a Poisson multi-Bernoulli density that is projected back to Poisson via minimizing the KLD.}
	\label{fig_KLD}
\end{figure}
\par In this paper, we propose a TPHD filter for scenarios in which there can be coexisting point and extended targets. {\color{black}The proposed TPHD filter combines the computational efficiency of PHD filters, with the ability to estimate trajectories from first principles in situations with both point and extended targets.} In particular, the contributions of this paper are:
\begin{enumerate}		
	\item A derivation of the TPHD filter update for general target-generated measurement density and PPP clutter based on direct KLD minimization (see Fig. \ref{fig_KLD}). The corresponding derivation for the general PHD filter was derived using probability generating functionals (PGFLs) \cite{ClarkGPHD2012, Mahler2007RFSbook}. Therefore, the proposed derivation increases the accessibility of the proof to a wider audience.
	\item An implementation of the general TPHD filter for tracking coexisting point and extended targets. The implementation is derived for a linear Gaussian model for point targets \cite{Vo2006PHD, Angel2019TPHD} and a Gamma Gaussian Inverse Wishart (GGIW) model for extended targets \cite{Grans2012exCPHD,ExPMBM2020}.
\end{enumerate}
\par The structure of this paper is organized as follows. In Section II, the theoretical background and recursion step of the general TPHD filter are provided. The implementation is given in Section III. In Section IV, the performance of the general TPHD filter, compared to other PHD filters, is evaluated via simulation and experimental data. Finally, conclusion are drawn in Section V,

\section{The General TPHD Filter}
In this section, the recursion steps of the general TPHD filter are provided. The main idea is to use a general likelihood for target-generated measurements in the update step. We define sets of trajectories, the Bayesian filtering recursion for sets of trajectories, and provide the update and prediction steps of the general TPHD filter. 
\subsection{Sets of Trajectories}
The trajectory state $X=(t,x^{1:i})$ consists of a finite sequence
of target states $x^{1:i}=(x^1,...,x^i)$ that starts at the time step $t$ with length $i$, {\color{black}where $x^i \in \mathbb{X}^i$ \cite{Angel2019TPHD}.} We use the notation $\mathbb{X}$ to denote the single-target space. For $k$ denoting the current time step and a trajectory $(t,x^{1:i})$ that exists from time step $t$ to $t+i-1$, the variable $(t,i)$ belongs to the set $\mathbb{I}_k=\{(t,i):1\leq t\leq k~\text{and}~1\leq i\leq k-t+1\}$. Therefore, a single trajectory up to the time step $k$ is defined in the space $\mathbb{T}_{k}=\uplus_{(t, i)\in \mathbb{I}_k}\{t\}\times\mathbb{X}^i$, where $\uplus$ denotes the disjoint union, $\times$ denotes a Cartesian product, and $\mathbb{X}^{i}$ represents the general target trajectory state space. Supposing there are $N^k$ trajectories at the time step $k$, the set of trajectories is denoted as 
\begin{align}
	\mathbf{X}_k=\{X_{1},...,X_{N^k}\}\in \mathcal{F}(\mathbb{T}_{k}),
\end{align} 
where $\mathcal{F}(\mathbb{T}_{k})$ represents the set of all finite subsets of $\mathbb{T}_{k}$. 
\subsection{Bayesian Filtering Recursion}
Given the posterior multi-trajectory density $\pi_{k-1}(\cdot)$ on the set of trajectories at the time step $k-1$ and the set of measurements $\mathbf{z}_k$ at the time step $k$, the posterior density $\pi_{k}(\cdot)$ is obtained by using the Bayes' recursion \cite{Angel2020TMTT}
\begin{align}
	\pi_{k|k-1}\left(\mathbf{X}_k \right) =& \int {\phi\left( {\mathbf{X}_k |\mathbf{X}_{k-1} } \right)} {\pi_{k - 1}}\left( \mathbf{X}_{k-1} \right)\delta \mathbf{X}_{k-1} , \\
	%\end{equation}
	%\begin{equation}
	\label{Bayes}\pi_k\left( \mathbf{X}_k \right) =& \frac{{{\ell_k}\left( {\mathbf{z}_k|\mathbf{X}_k} \right){\pi_{k|k - 1}}\left( \mathbf{X}_k \right)}}{	\int {{\ell_k}\left( {\mathbf{z}_k| \mathbf{X}_k} \right)} {\pi_{k|k - 1}}\left( \mathbf{X}_k \right)\delta \mathbf{X}_k},
\end{align}
where $\phi( { \cdot | \cdot })$ denotes the transition density for sets of trajectories, $ \pi_{k|k - 1}(\cdot)$ denotes the predicted density, ${\ell_k}\left(\mathbf{z}_k|\mathbf{X}\right)$ denotes the density of measurements of trajectories. As the measurements $\mathbf{z}_k$ come from the target states at the current time step $k$, ${\ell_k}\left( \mathbf{z}_k|\mathbf{X}_k \right)$ can be also written as
\begin{equation}
	{\ell_k}\left( {\mathbf{z}_k|\mathbf{X}_k} \right) = {\ell_k}\left( {\mathbf{z}_k|{\tau _k}\left( \mathbf{X}_k \right)} \right),
\end{equation}
where $\tau_k(\mathbf X)$ denotes the corresponding multi-target state at the time step $k$. 
\par We use the MMSE estimation to obtain the trajectory state \cite{Angel2019TPHD, Vo2006PHD}. It is worth to mention that the current state estimation performance of TPHD filters are the same as for the PHD filters. That is, a TPHD filter enables the estimation of past states of the trajectories, but
a TPHD filter has the same information regarding the current state of the trajectories as a PHD filter.

\subsection{Update}
The general TPHD filter propagates a Poisson density on the set of alive trajectories through the filtering recursion via KLD minimization \cite{Angel2015KLD}, see Fig.\ref{fig_KLD}. Given the current set of targets, the measurement model is:
\par \emph{Assumption 1:} Each target generates an independent set $\mathbf{z}_k$ of measurements with density $f\left(\mathbf{z}_k|\cdot\right)$. 
\par \emph{Assumption 2:} The set $\mathbf{z}_{k}$ of measurements at time step $k$ is the union of target generated measurements and clutter. Clutter is a PPP with intensity $\lambda^{C}\left(\cdot\right)$, which is independent of target-originated measurements. 
\par Let $\lambda_{k|k}\left(X\right)$ be the PHD of the alive trajectories $X$ at time step $k$. For $X=(t,x^{1:i})$, we use $\lambda_{k|k}\left(x^i\right)$ as the PHD of the current set of targets, which can be obtained by marginalizing the PHD for trajectories \cite{Angel2019TPHD}. For the general TPHD filter, we use a pseudolikelihood function $L_{\mathbf{z}_{k}}(\cdot)$ for the update step \cite{Mahler2014RFSbook, ClarkGPHD2012}, which is given as
\begin{align}\label{likelihood}
	L_{\mathbf{z}_{k}}\left(x^i\right) & =f\left(\emptyset|x^i\right)+\sum_{\mathcal{P}\angle\mathbf{z}_{k}}w_{\mathcal{P}}\sum_{\mathbf{w}\in\mathcal{P}}\frac{f\left(\mathbf{w}|x^i\right)}{\kappa_{\mathbf{w}}+\tau_{\mathbf{w}}}
\end{align}
where 
\begin{align}
	\tau_{\mathbf{w}} & =\int f\left(\mathbf{w}|x^i\right)\lambda_{k|k-1}\left(x^i\right)dx^i\label{eq:tau_w},\\
	%\end{align}
	%\begin{align}
	\kappa_{\mathbf{w}} & =\delta_{1}\left[|\mathbf{w}|\right]\left[\prod_{z\in\mathbf{w}}\lambda^{C}\left(z\right)\right],\quad|\mathbf{w}|>0,\\
	w_{\mathcal{P}} & =\frac{\prod_{\mathbf{w}\in\mathcal{P}}\left(\kappa_{\mathbf{w}}+\tau_{\mathbf{w}}\right)}{\sum_{\mathcal{Q}\angle\mathbf{z}_{k}}\prod_{\mathbf{w}\in\mathcal{Q}}\left(\kappa_{\mathbf{w}}+\tau_{\mathbf{w}}\right)}.\label{eq:w_P}
\end{align}
{\color{black} In \eqref{likelihood}, the notation $\mathcal{P}\angle\mathbf{z}_{k}$ denotes that the sum goes over all partitions $\mathcal{P}$ of $\mathbf{z}_{k}$. Then, the sum $\mathbf{w}\in\mathcal{P}$ goes through all sets in this partition \cite{Grans2012exPHD1, GransXradar2015}. We use $w_{\mathcal{P}}$ to denote the weight of each partition \cite{Mahler2014RFSbook}. The notation $\delta_i[\cdot]$ is a Kronecker delta located at $i$.}

\begin{Pro} 
	Given the prior trajectory PHD with intensity $\lambda_{k|k-1}\left(t,x^{1:i}\right)$ at the current time step $k$, the TPHD filter update is
	\begin{align}
		\lambda_{k|k}\left(t,x^{1:i}\right) & =L_{\mathbf{z}_{k}}\left(x^{i}\right)\lambda_{k|k-1}\left(t,x^{1:i}\right),\label{eq:TPHD_update}
	\end{align}
	if $t+i-1=k$ and otherwise $\lambda_{k|k-1}\left(t,x^{1:i}\right)$ equals to zero.
\end{Pro}
\par {\color{black}The proof of Proposition 1 via direct KLD minimization is provided in Appendix A. As a general target-generated measurement model is considered in this filter, we can recover the standard point and extended TPHD filter updates (See Appendix B). The general TPHD (G-TPHD) filter is not a straightforward combination of the TPHD filter for point target (P-TPHD) and TPHD for extended target tracking (E-TPHD). To derive the G-TPHD, we first derive the TPHD update for general target-generated measurement, which can take into account the measurement models for point and extended targets, simultaneously. Without this measurement model and the associated update step, it is not possible to derive the G-TPHD filter.}

\subsection{Prediction}
The general TPHD filter considers the standard dynamic models, and therefore, the prediction step is similar to the standard TPHD filter \cite{Angel2019TPHD}.
Given the current set of targets, the multi-target dynamic model is:
\par\emph{Assumption 3:} A target $x$ survives to the next time step with probability $p^S(x)$ and transition density $g(\cdot|x)$.
\par\emph{Assumption 4:} New targets are born independently following a PPP with intensity $\lambda_{\gamma}(\cdot)$. The set of targets at the next time step is the union of surviving targets and new born targets.
\begin{Pro}
	Given the posterior trajectory PHD $\lambda_{k - 1|k-1}\left( t,x^{1:i - 1} \right)$ at the last time step $k-1$ and birth PHD at the current time step $k$, the prediction of the general TPHD filter is \cite{Angel2019TPHD}
	\begin{align}\label{equ_TPHD_pr_total}
		{\lambda_{k|k - 1}}\left( X \right)= {\lambda_{\gamma,k|k}}\left( X \right) + \lambda_{k|k-1}^S \left( X \right),
	\end{align}
	where
	\begin{align}
		\label{eq_birth}\lambda_{\gamma,k|k}\left( X \right) =& \lambda_{\gamma} \left( x^1 \right)\delta_{1}[i]\delta_{k}[t],\\
		\label{eq_survive}\lambda_{k|k-1}^S \left( X \right) =&  {p^{S}}\left( {{x^{i-1}}} \right)
		g\left( {x^i|x^{i-1}} \right){\lambda_{k - 1|k-1}}\left( t,x^{1:i - 1} \right).
	\end{align}
\end{Pro}
\par Eq. \eqref{equ_TPHD_pr_total} is the sum of the intensities for new born trajectories \eqref{eq_birth} and surviving trajectories \eqref{eq_survive}.  

\section{The Gamma Gaussian Inverse Wishart Implementation}
In this section, we apply the general TPHD filter recursion in Section II to track coexisting point extended targets. First, we explain the space and then the implementations. Finally, some strategies are given to decrease the computational cost. %filter to track both point and extended targets using the Gamma Gaussian Inverse Wishart (GGIW) mixtures. The PHD of the coexisting point and extended target trajectory $\lambda_{k|k}(X)$ at time $k$ can be also written as, $\lambda_{k|k}(X)=\lambda_{k|k}(X_p)+\lambda_{k|k}(X_e)$. 
\begin{figure*}[!t]
	\centering
	\includegraphics[width=4.5in]{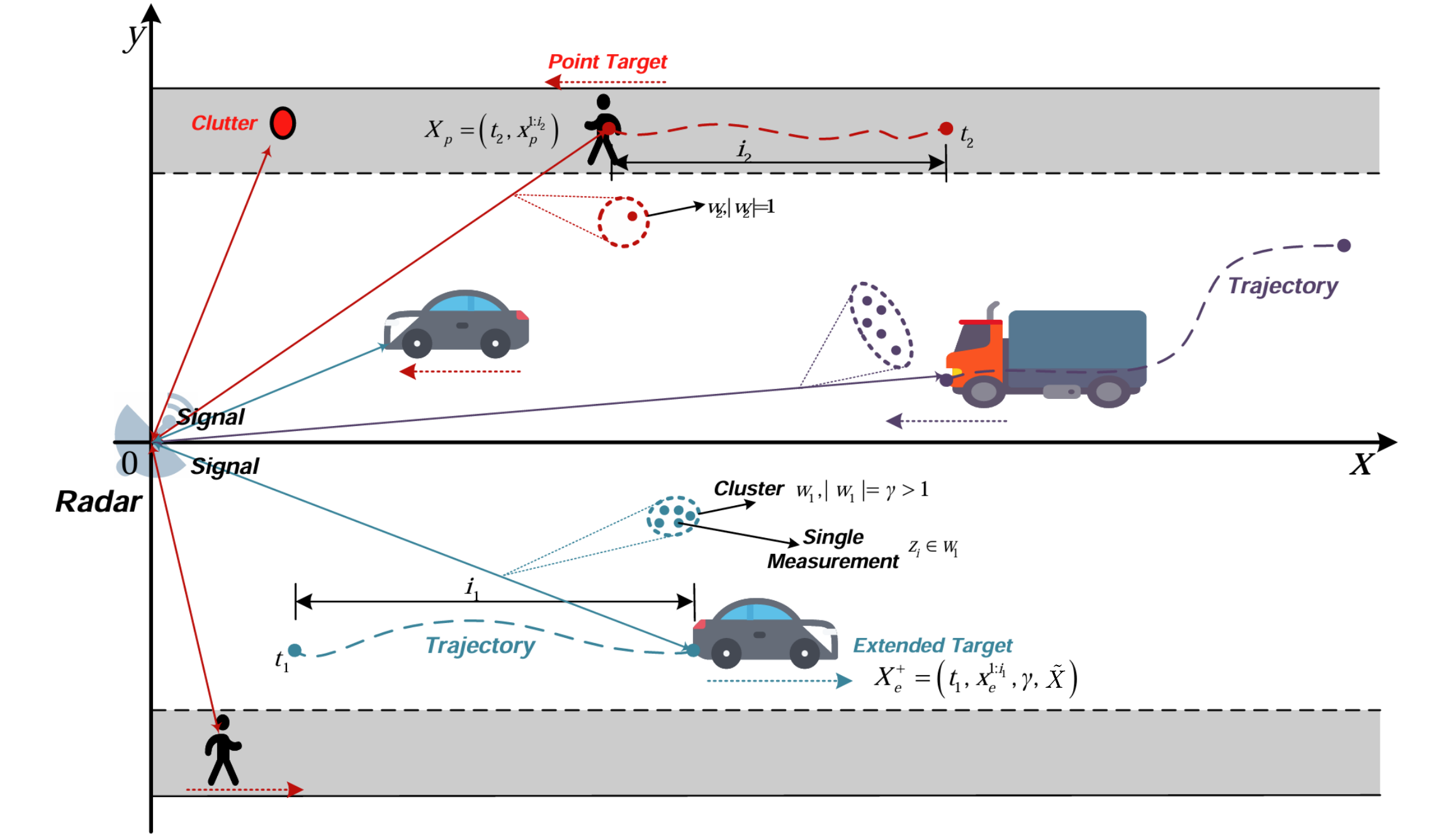}
	\caption{This figure shows a scenario consisting of both point and extended targets. For a given radar resolutions, pedestrians can be considered as point targets and vehicles as extended targets (as in the experimental results section). The measurements of each potential target form a cluster. The point target will generate at most one measurement and the extended target will generate a time-varying number of measurements within a certain range and angular spread per time step.}
	\label{traffic}
\end{figure*}
\subsection{The Coexisting Point-Extended Target Space Model}
A common scenario for coexisting extended and point target is a traffic monitoring situation, as illustrated in Fig. \ref{traffic}. To track both kinds of targets, it is important to define a general space model. First, for the scenario in Fig. \ref{traffic}, we define the state of point target trajectory at the time step $k$ with the notation:
\begin{align}
	{X_p}=(t,x_p^{1:i}),
\end{align}
where $t$ represents its birth time and $x_p^{1:i}=(x_p^1,...,x_p^i)$ denotes a sequence including the point target states at each time step of the trajectory with length $i$ \cite{Angel2019TPHD}. The state $X_p$ belongs to the space $\mathbb{T}_{p,k}$, which is written as
\begin{align}
	{X_p}\in\mathbb{T}_{p,k}=\uplus_{(t,i)\in \mathbb{I}_k}\{t\}\times\mathbb{R}^{i\times n_x},
\end{align}
\par To describe an extended target state $X^+_e$ at the time step $k$, we respectively define the notation $X_e$ for trajectory state, $\gamma$ for the expected number of measurements per target and $\tilde X$ for the extent state\cite{Extracking-overview2017}. More specifically,
\begin{align}
	X^+_e=(X_e,\gamma,\tilde X)\in&\mathbb{T}_{e,k},\\
	X_e=(t,x^{1:i}_e)\in&\uplus_{(t,i)\in \mathbb{I}_k}\{t\}\times\mathbb{R}^{i\times n_x},\\
	\gamma\in& \mathbb{R}_{+},\\
	\tilde X \in&\mathbb{S}^d_+,
\end{align}
where $x_e^{1:i}$ denotes the sequence of target states, $\mathbb{R}_{+}$ represents the space of positive real numbers and $\mathbb{S}^d_+$ is the space of positive definite matrices with size $d$. In other words, the state space for extended targets can be written as $\mathbb{T}_{e,k} = \uplus_{(t,i)\in \mathbb{I}_k}\{t\}\times\mathbb{R}^{i\times n_x}\times \mathbb{R}_{+}\times\mathbb{S}^d_+ $. The value of $d$ is taken as the dimension of the target extent. Here, we only consider target extent $\tilde X$ and $\gamma$ at the latest time step for simplicity. If the historical information needs to be stored, we just add the corresponding sequence to the trajectory space. Therefore, the coexisting trajectory space for both point and extended targets is $\mathbb{T}_k = \mathbb{T}_{p,k}\uplus \mathbb{T}_{e,k}$.

\subsection{The GGIW Model}
\par The implementation is provided for a Gaussian model for point target trajectory \cite{Filtering-2013, Vo2006PHD, Angel2019TPHD} and a Gamma Gaussian Inverse Wishart (GGIW) model for extended target trajectory. \cite{Grans2012exCPHD,ExPMBM2020,Extracking-overview2017,GransXradar2015}.
\par The density of $\gamma$ is given as the Gamma distribution ${\cal G}(\gamma;a,b)$ with parameters $a>0$ and $b>0$. The Inverse Wishart density on matrices ${\cal{IW}}(\tilde X;v,V)$ is used to describe $\tilde X$, which is defined in space $\mathbb{S}^d_+$ with $v > 2d$ degrees of freedom and parameter matrix $V\in\mathbb{S}^d_+$. The Gamma and Inverse Wishart distributions are  conjugate priors to the Poisson distribution and multivariate Gaussian distribution, respectively \cite{Exgamma,Exwishart}.
\par At the time step $k$, the Gaussian density of the trajectory $X=(t,x^{1:i})$ is denoted as~\cite{Angel2019TPHD}
\begin{align}\label{Gaussian}
	{\cal N}( {X;t^k,\widehat{m}^k,\widehat{P}^k})={\cal N}( {x^{1:i};\widehat{m}^k,\widehat{P}^k})\delta_{[t^k]}[t]\delta_{[i^k]}[i].
\end{align}
where $\widehat{m}^k \in \mathbb{R}^{i{n_x}}$ and $\widehat{P}^k \in \mathbb{R}^{i{n_x}\times i{n_x}}$ denote the mean and covariance, respectively. The term $t^k=k-i^k+1$ denotes the trajectory birth time and $i^k=\text{dim}(\widehat{m}^k/n_x)$ is the trajectory length.
\par Then, the PHD $\lambda_{k|k}(X)$ at the time step $k$ is
\begin{align}\label{De_PHD}
	\lambda_{k|k}(X)=\lambda_{k|k}^e(X)+\lambda_{k|k}^p(X),
\end{align}
where $\lambda_{k|k}^e(X)=0$ if $X \in \mathbb{T}_{p,k}$ and $\lambda_{k|k}^p(X)=0$ if $X \in \mathbb{T}_{e,k}$. Therefore, we can write the two terms at the right side of the above equation as
\begin{align}
	\label{w_e}{\lambda^e_{k|k}}\left( X^+_e \right)=&\sum\limits_{j = 1}^{{J_e^{k}}} \omega_{e,j}^{k}{\cal G}\left( {\gamma;a_j^{k},b_j^{k} } \right)\notag\\
	&\times{\cal N}\left( {X_e;t_{e,j},\widehat m_{e,j}^{k},\widehat P_{e,j}^{k}} \right)\notag\\
	%\end{align}
	%\begin{align}
	&\times{\cal{IW}}\left( {\tilde X;v_j^{k},V_j^{k} } \right),\\
	\label{w_p}{\lambda^p_{k|k}}\left( X_p \right)=&\sum\limits_{i = 1}^{{J_p^{k}}} \omega_{p,i}^{k}{\cal N}\left( {X_p;t_{p,i},\widehat m_{p,i}^{k},\widehat P_{p,i}^{k}} \right),	
\end{align}
where $J_e^{k}, J_p^{k}$ denote the number of GGIW and Gaussian components. The notations $w_{e,j}^{k}, w_{p,i}^{k}$ denote the weight for each GGIW and Gaussian component.
\par {\color{black} We know that a target \( x \in \mathbb{R}^{n_x} \uplus \mathbb{E} \), where the extended target space $\mathbb{E} = \mathbb{R}^{n_x} \times \mathbb{R}_{+}\times \mathbb{S}^{d}_{+}$. The transition density to convert a point target into an extended target is \( g_{p,e}(\cdot|\cdot) \), and the transition density to convert an extended target into a point target is \( g_{e,p}(\cdot|\cdot) \). Then, the overall transition density is
\begin{align}
		&g(x^i|x^{i-1}) = \notag\\
	&\begin{cases} 
		s_p (x^{i-1}) g_p(x^i|x^{i-1}) & \text{if } x^{i-1} \in \mathbb{R}^{n_x} \\
		(1- s_p (x^{i-1})) g_{p,e}(x^i|x^{i-1}) & \text{if } x^{i-1} \in \mathbb{R}^{n_x} \\
		s_e (x^{i-1}) g_e(x^i|x^{i-1}) & \text{if } x^{i-1} \in \mathbb{E}\\
		(1- s_e (x^{i-1})) g_{e,p}(x^i|x^{i-1}) & \text{if } x^{i-1} \in \mathbb{E}
	\end{cases}
\end{align}
where \( s_p (x^{i-1}) \) is the probability that a point target with state \( x_{i-1} \) stays as a point target and \( s_e (x^{i-1}) \) is the probability
that an extended target with state \( x^{i-1} \) stays as an extended target. We obtain
closed-form formulas by setting \( s_p (x^{i-1}) = s_p \) and \( s_e (x^{i-1}) = s_e \).}  Besides, the surviving and detection probability is also constant in implementation, i.e., $p^S(x)=p^S, p^D(x)=p^D$. The parameter $\gamma$ is also assumed to be constant.

\par {\color{black}The measurement density given a single target $f(\mathbf{z}_k|\cdot)$ in Assumption 1 can cover the measurements from both point and extended targets. If the target $x$ is a point target, it is detected with probability $p^D$ and generates a single-measurement with likelihood $l(z|x^i)={\cal N}(z;Hx_p^{i},R)$, where $z\in \mathbf{z}_k$ denotes a single measurement. If the target is an extended target, it is detected with probability $p^D$, and, if detected, it can generate multiple measurements, modeled by a Poisson point process. The single-measurement likelihood is $l(z|x^i)={\cal N}(z;Hx_e^{i},\tilde X)$. The specific forms of $f(\mathbf{z}_k|\cdot)$ for point and extended targets are explained in Appendix B. }

\par For simplicity in following calculations, we define that, for a matrix $V$, the notation ${V_{\left[ {n:m,s:t} \right]}}$ represents the submatrix of $V$ for rows from time steps $n$ to $m$ and columns from time steps $s$ to $t$.

\subsection{Update}
The update step is given by the following proposition.
\begin{Pro}\label{TPHD_Up}
	Given the prior PHD $\lambda_{k|k-1}(\cdot)$ of the form \eqref{De_PHD} and the measurement set $\mathbf{z}_k$, the posterior PHD $\lambda_{k|k}(\cdot)$ is
	\begin{align}
		{\lambda_{k|k}}\left( X \right)={\lambda_{k|k}^{ND}}\left( X \right)+\sum_{\mathcal{P}\angle \mathbf{z}_k}\sum_{\mathbf{w}\in\mathcal{P}}{\lambda_{k|k}^{D}}\left( X,\mathbf{w} \right).
	\end{align}
\end{Pro}
Following \eqref{De_PHD}, the mis-detected part ${\lambda_{k|k}^{ND}}\left( X \right)$ can be written as the sum of $\lambda_{k|k}^{e,ND}(X)$ and $\lambda_{k|k}^{p,ND}(X)$, which are respectively given as
\begin{align}
	{\lambda_{k|k}^{e,ND}}\left( X^+_e \right)=&\sum\limits_{j = 1}^{{J_e^{k|k - 1}}}\omega_{e,j}^{k|k-1}{\cal N}\left( {X_e;t_{e,j},\widehat m_{e,j}^{k|k - 1},\widehat P_{e,j}^{k|k - 1}} \right)\notag\\
	&\times{\cal{IW}}\left( {\tilde X;v_j^{k|k - 1},V_j^{k|k - 1} } \right)\notag\\
	&\times\left[(1-p^D) {\cal G}\left( {\gamma;a_j^{k|k - 1},b_j^{k|k - 1} } \right)\right. \notag\\
	&\left.+p^D\left(\frac{b_j^{k|k-1}}{b_j^{k|k-1}+1}\right)^{a_j^{k|k-1}}\right.\notag\\
	&\left. \times{\cal G}\left( {\gamma;a_j^{k|k - 1},b_j^{k|k - 1}+1 } \right)\right],
 \end{align}
\begin{align} 
	{\lambda_{k|k}^{p,ND}}\left( X_p \right)=&\sum\limits_{i = 1}^{{J_p^{k|k - 1}}} (1-p^D) \omega_{p,i}^{k|k-1}\notag\\
	&\times{\cal N}\left( {X_p;t_{p,i},\widehat m_{p,i}^{k|k - 1},\widehat P_{p,i}^{k|k - 1}} \right).
\end{align}
\par For extended undetected targets, each component of the prior intensity mixture gives rise to two components in the update step. The first corresponds to the situation of the missed detection, which is modeled by $p^D$. The second corresponds to the situation that the Poisson random number of detections is zero, also governed by the factors $a,b$ of Gamma density \cite{ExPMBM2020,Exgamma}. For undetected point targets, each component in the prior intensity mixture gives rise to only one component, corresponding to misdetection \cite{Vo2006PHD,Angel2019TPHD}. Similarly, the detected part ${\lambda_{k|k}^{D}}(X,\mathbf{w})$ can be decomposed as
\begin{align}\label{Up_ep}
	{\lambda_{k|k}^{e,D}}\left( X^+_e,\mathbf{w} \right)=&\sum\limits_{j = 1}^{{J_e^{k|k - 1}}} \omega_{e,j}^{k}(\mathbf{w}){\cal G}\left( {\gamma;a_j^{k},b_j^{k} } \right)\\
	\times&{\cal N}\left( {X_e;t_{e,j},\widehat m_{e,j}^{k},\widehat P_{e,j}^{k}} \right){\cal{IW}}\left( {\tilde X;v_j^{k},V_j^{k} } \right),\notag\\
	%\end{align}
	%\begin{align}
	{\lambda_{k|k}^{p,D}}\left( X_p,\mathbf{w} \right)=&\sum\limits_{i = 1}^{{J_p^{k|k - 1}}}\omega_{p,i}^{k}(\mathbf{w}){\cal N}\left( {X_p;t_{p,i},\widehat m_{p,i}^{k},\widehat P_{p,i}^{k}} \right),
\end{align}
where
\begin{align}
	\omega_{e,j}^{k}(\mathbf{w}) =& \frac{\omega_{\mathcal{P}}}{d_{\mathbf{w}}}\omega_{e,j}^{k|k - 1}p^D\cdot\frac{q_{e,j}(\mathbf{w})}{\left[\lambda^{C}\left(\cdot\right)\right]^{\mathbf{w}}},\label{w_extend}\\
	\omega_{p,i}^{k}(\mathbf{w}) =& \frac{\omega_{\mathcal{P}}}{d_{\mathbf{w}}}\omega_{p,i}^{k|k - 1}p^D\cdot \frac{q_{p,i}(\mathbf{w})}{\lambda^{C}\left(\cdot\right)}\delta_{1}[\vert \mathbf{w} \vert],\label{w_point}\\
	\omega_{\mathcal{P}} = &\frac{\prod_{\mathbf{w}\in\mathcal{P}}d_{\mathbf{w}}}{\sum_{\mathcal{Q}\angle\mathbf{z}_{k}}\prod_{\mathbf{w}\in\mathcal{Q}}d_{\mathbf{w}}},\\
	%\end{align}
	%\begin{align}
	d_{\mathbf{w}}=&\delta_{1}[\vert \mathbf{w} \vert]+ \sum\limits_{l = 1}^{{J_e^{k|k - 1}}} \omega_{e,l}^{k|k - 1}p^D\frac{q_{e,l}(\mathbf{w})}{\left[\lambda^{C}\left(\cdot\right)\right]^{\mathbf{w}}}\notag\\
	&+ \sum\limits_{o = 1}^{{J_p^{k|k - 1}}} {\omega_{p,o}^{k|k - 1}p^D \frac{q_{p,o}(\mathbf{w})}{\lambda^{C}\left(\cdot\right)}\delta_{1}[\vert \mathbf{w} \vert]}.\label{dw_point}
\end{align}
\par It is worth noting that if $\vert \mathbf{w} \vert > 1$, the target must be an extended target, but
if $\vert \mathbf{w} \vert = 1$, the target may be a point, an extended target or the clutter. Therefore, we add the delta function $\delta_{1}[\vert \mathbf{w} \vert]$ in \eqref{w_point} and \eqref{dw_point}. {\color{black} From \eqref{w_extend}--\eqref{dw_point}, we can see that the intensities of trajectories of point and extended targets are obtained according to the prior and the weights of the different data-association hypotheses of the G-TPHD filter recursion. In particular, the weights of the intensity components for point targets, depend on that of the extended targets, and viceversa.} 
\par For the specific update for the factors of GGIW components in \eqref{Up_ep}, we have following equations:
\begin{align}
	&(\widehat{m}_{e,j}^k,\widehat{P}_{e,j}^k,a^k_j,b^k_j,v^k_j,V^k_j)\notag\\
	%\end{align}
	%\begin{align}
	&=	\left\{
	\begin{array}{ll}
		\widehat{m}_{e,j}^k&= \widehat{m}_{e,j}^{k|k - 1} + {K_j}\epsilon_j,\\
		\widehat{P}_{e,j}^k&= \widehat{P}_{e,j}^{k|k - 1} -{K_j}H\widehat{P}_{{e,j},[ {k,t_j:k}]}^{k|k - 1},\\
		a_j^k&=a_j^{k|k-1}+\vert \mathbf{w} \vert,\\
		b_j^k&=b_j^{k|k-1}+1,\\
		v_j^k&= v_j^{k|k - 1} +\vert \mathbf{w} \vert,\\
		V_j^k&= V_j^{k|k - 1} + N_j+Z,\\
	\end{array}
	\right.
\end{align}
where
\begin{align}
	{\bar z} =& \frac{1}{\vert \mathbf{w} \vert}\sum_{z\in \mathbf{w}}z,\\
	Z =&\sum_{z\in \mathbf{w}}(z-\bar z)(z-\bar z)^\top,\\
%\end{align}
%\begin{align}
	\tilde X_j =&\frac{V_j^{k|k-1}}{v_j^{k|k-1}-2d-2},\\
	\epsilon_j =& \bar{z}-{H}\widehat m_{{e,j},[k]}^{k|k - 1},\\
	{S_j} =& {H}\widehat P^{k|k-1}_{{e,j},[k,k]}H^\top + \frac{\tilde X_j}{\vert \mathbf{w} \vert},\\
	{K_j} =& \widehat{P} _{{e,j},[ {t_j:k,k} ]}^{k|k - 1}H^\top S_j^{ - 1},\\
	N_j = &\tilde X_j^{{1}/{2}}S_j^{-1/{2}}\epsilon_j\epsilon_j^\top S_j^{-{\top}/{2}}\tilde X_j^{\top/2}.
\end{align}
\par To calculate the direct Gaussian update for point targets, i.e., $(\widehat{m}_{p,j}^k,\widehat{P}_{p,j}^k)$, we follow the same principles as \cite{Angel2019TPHD}. Besides, the measurement likelihood function for coexisting point and extended targets is obtained by:
\begin{align}
	{q_{p,j}}\left( \mathbf{w} \right) =& {\cal N}\left( {\mathbf{w};{H}\widehat{m}_{{p,j},[k]}^{k|k - 1},{H}\widehat P^{k|k-1}_{{p,j},[k,k]}H^\top + R} \right),\\
	%\end{align}
	%\begin{align}
	{q_{e,j}}\left( \mathbf{w} \right) =& \left( \pi^{\vert \mathbf{w} \vert} \vert \mathbf{w} \vert \right)^{-d/2}\frac{\left\vert V_j^{k|k-1} \right \vert^{\frac{v_j^{k|k-1}-d-1}{2}}}{\left\vert V_j^{k} \right\vert^{\frac{v_j^{k}-d-1}{2}}}\\
	&\times\frac{\Gamma_d\left(\frac{v_j^{k}-d-1}{2}\right)\left\vert \tilde X_j \right\vert^{1/2}\Gamma(a_j^k)(b_j^{k|k-1})^{a_j^{k|k-1}}}{\Gamma_d\left(\frac{v_j^{k|k-1}-d-1}{2}\right)\vert S_j \vert^{1/2}\Gamma(a_j^{k|k-1})(b_j^{k})^{a_j^{k}}},\notag
\end{align}
where $\Gamma$ is the Gamma function and $\Gamma_d$ is the multivariate Gamma function. Then, we complete the process of update step in general TPHD filter using GGIW mixture. The estimation step is given in Appendix E.
\par {\color{black}Generally, if the targets have been born far away and have been classified with high certainty as point or extended targets, the filter would be able to distinguish the type of nearby targets using the dynamic model (assuming that changes of target type happen with low probability). However, if we receive an isolated cluster of measurements in which the birth process enables the appearance of point and extended targets, it is not possible to determine with certainty if there is an extended target or multiple point targets.  After several time steps have passed and the targets separate, the filter should be able to classify the type of each target well.}
\subsection{Prediction}
{\color{black}The prediction step is given by the following proposition. We commence by introducing the intensity of new born targets, which is given as
\begin{align}
	{\lambda_{\gamma,k|k}}\left( X\right) = {\lambda^e_{\gamma,k|k}}\left( X\right)+{\lambda^p_{\gamma,k|k}}\left( X\right)
\end{align}
where
\begin{align}
	{\lambda^e_{\gamma,k|k}}\left( X^+_e\right) =& \sum\limits_{j = 1}^{J_{\gamma^e}^{k}}{\omega _{\gamma^e,j}^{k} {\cal G} \left( {\gamma;a_{\gamma,j}^{k},b_{\gamma,j}^{k}} \right)}\\
	&\times{{\cal N}\left( {X_e;k,\widehat{m} _{\gamma^e,j}^{k},\widehat{P} _{\gamma^e,j}^k} \right)}\notag\\
	&\times{\cal{IW}}\left( {\tilde X;v_{\gamma,j}^{k},V_{\gamma,j}^{k} } \right),\notag\\
	%\end{align}
	%\begin{align}
	{\lambda^p_{\gamma,k|k}}\left( X_p\right) =& \sum\limits_{i = 1}^{{J_{\gamma^p}^{k }}}\omega _{\gamma^p,i}^{k} {{\cal N}\left( {X_p;k,\widehat{m} _{\gamma^p,i}^{k},\widehat{P} _{\gamma^p,i}^{k}} \right)}.
\end{align}
The notations $J_{\gamma^e}^{k}$ and $J_{\gamma^p}^{k}$ denote the number of PHD components for new born targets. For the $j$-{th} birth component at the time step $k$, $\omega_{\gamma^e,j}^k$ and $\omega_{\gamma^e,j}^k$ represent the weight of extended and point target, respectively. 
%\par \emph{Assumption 10:} The born time $t_ j$ of a single trajectory $X_j$ is considered as a constant.
\begin{Pro}\label{BGUTPHD_Pr}
	Given the posterior PHD $\lambda_{k-1|k-1}(\cdot)$ for coexisting point and
	extended target trajectories at the time step $k-1$, the prior PHD $\lambda_{k|k-1}(\cdot)$ is
	\begin{align}
		{\lambda_{k|k - 1}}\left( X \right) =&\lambda_{\gamma,k|k}\left( X \right)+\lambda^S_{k|k-1}\left( X \right),\\
		\label{four_term_pre}\lambda^S_{k|k-1}\left( X \right)=&\lambda^{S,e}_{k|k-1}\left( X \right)+\lambda^{S,p}_{k|k-1}\left( X \right)\\
		&+ \lambda^{S,p,e}_{k|k-1}(X) + \lambda^{S,e,p}_{k|k-1}(X)\notag,
	\end{align}
	where
	\begin{align}
		{\lambda^{S,e}_{k|k - 1}}\left( X^+_e \right) =&s_e\cdot{p^{S}}\sum\limits_{j = 1}^{{J_e^{k - 1}}} {{\omega} _{e,j}^{k - 1}{\cal G} \left( {\gamma;a_{S,j}^{k|k - 1},b_{S,j}^{k|k - 1}} \right)}\notag\\
		&\times{{\cal N}\left( {X_e;t_{e,j},\widehat{m} _{S^e,j}^{k|k-1},\widehat{P} _{S^e,j}^{k|k - 1}} \right)}\notag\\
		&\times {\cal{IW}}\left( {\tilde X;v_{S,j}^{k|k - 1},V_{S,j}^{k|k - 1} } \right),\\
		{\lambda^{S,p}_{k|k - 1}}\left( X_p \right) =& s_p\cdot p^{S} \sum\limits_{i = 1}^{{J_p^{k - 1}}}{\omega} _{p,i}^{k - 1}\notag\\
		&\times{{\cal N}\left( {X_p;t_{p,i},\widehat{m} _{S^p,i}^{k|k-1},\widehat{P} _{S^p,i}^{k|k - 1}} \right)}.\\
  		\lambda^{S,p,e}_{k|k-1}(X^+_e) =&(1-s_p){p^{S}}\sum\limits_{i = 1}^{{J_p^{k - 1}}} {{\omega} _{p,i}^{k - 1}{\cal G} \left( {\gamma;a_{B},b_{B}} \right)}\notag\\
		&\times{{\cal N}\left( {X_e;t_{p,i},\widehat{m} _{S^p,i}^{k|k-1},\widehat{P} _{S^p,i}^{k|k - 1}} \right)}\notag\\
		&\times {\cal{IW}}\left( {\tilde X;v_{B},V_{B}} \right),
  \end{align}
\begin{align}
		{\lambda^{S,e,p}_{k|k - 1}}\left( X_p \right) =& s_p\cdot p^{S} \sum\limits_{j = 1}^{{J_e^{k - 1}}}{\omega} _{e,j}^{k - 1}\notag\\
		&\times{{\cal N}\left( {X_p;t_{e,j},\widehat{m} _{S^e,j}^{k|k-1},\widehat{P} _{S^e,j}^{k|k - 1}} \right)}.
	\end{align}
\end{Pro}
In brief, the prediction of both point and extended target trajectory uses the same equation. To switch extended targets to point targets, we drop the information representing target extent. To switch point targets to extended targets, we consider a prior for the extend with parameters $a_B, b_B, v_B, V_B$. }
\par The predicted mean and covariance for surviving trajectories are:
\begin{align}
	\label{pr_begin}	\widehat m_{S^u,j}^{k|k - 1} =& \left[ {\left[\widehat m_{u,j}^{k - 1}\right]^\top,\left[F \cdot \widehat{m}_{{u,j},[k-1]}^{k - 1}\right]^\top} \right]^\top,\\
	\widehat P_{S^u,j}^{k|k - 1} =& \left[ {\begin{array}{*{20}{c}}
			{\widehat P_{u,j}^{k - 1}}&P_1\\
			P_1^\top&P_2
	\end{array}} \right],\\
	%\end{align}
	%\begin{align}
	P_1=&{\widehat P_{{u,j},\left[ {t_{u,j}:k - 1,k - 1} \right]}^{k - 1}{F^\top}},\\
	\label{pr_end}	P_2=&{F\widehat{P}_{{u,j},[k-1,k-1]}^{k - 1}F^\top + Q},
\end{align}
 where $F$ is the transition matrix and $Q$ is the process noise covariance matrix. It is assumed to follow the linear Gaussian model. We have $u = p$ or $u = e$. Then, the prediction steps for the Gamma and Inverse Wishart components are given as:
\begin{align}		
	a_{S,j}^{k|k-1}=&a_j^{k-1}/\mu,\label{mu}\\
	b_{S,j}^{k|k-1}=&b_j^{k-1}/\mu,\\
	v_{S,j}^{k|k-1}=&2d+2+e^{-\delta_t/\tau}(v_j^{k-1}-2d-2),\label{tau}\\
	V_{S,j}^{k|k-1}=&e^{-\delta_t/\tau}M(\widehat{m}_{{e,j},[k-1]}^{k - 1})V_j^{k-1}M^\top(\widehat{m}_{{e,j},[k-1]}^{k - 1}),\label{M}
\end{align}
where $\mu$ denotes the measurement rate parameter, $\delta_t$ is the sampling time period, $\tau$ is the correlation constant and $M(\cdot)$ is the transformation matrix for the extent model \cite{ExPMBM2020}.

\subsection{Strategies to Lower the Computational Cost}
In this subsection, we introduce three strategies to make filter implementation tractable. To restrict the unbounded increasing GGIW and Gaussian components, we can adopt the pruning and absorption techniques \cite{Angel2019TPHD}. 
\begin{table}[!t]
	\centering
	\caption{Pruning and absorption algorithm for TPHD filter with point and extended targets}
	\label{Algorithm}
	\begin{tabular}{p{8cm}}
		\toprule
		\midrule
		\textbf{Input:} Posterior PHD parameters~$\{\Phi_{e,j}^k\}_{j=1}^{J_e^k}$ for extended and $\{\Phi_{p,j}^k\}_{j=1}^{J_p^k}$ for point target trajectory, which are $$\Phi_{e,j}^k=\{\omega_{e,j}^k,t_{e,j},i_{e,j}^k,\widehat m_{e,j}^k,\widehat P_{e,j}^k,a^k_j,b^k_j,v^k_j,V^k_j \}$$
		and $$\Phi_{p,j}^k=\{\omega_{p,j}^k,t_{p,j},i_{p,j}^k,\widehat m_{p,j}^k,\widehat P_{p,j}^k \},$$		
		and the pruning threshold $\mathit{\Gamma_p}$, absorption threshold $\mathit{\Gamma_a}$ and maximum allowable number of components $J_{max}$.\\
		\textbf{For} $u=e$ or $p$:\\
		\textbf{Set} $\ell=0$ and $\Theta=\left\{i=1,...,J_u^k|\omega_{u,i}^k>\mathit{\Gamma_p}\right\}$.\\
		%\midrule
		\textbf{Loop}\\
		\par ~~$\ell=\ell+1.$\vspace{1mm}
		\par ~~$j$\,=\,\text{arg}${\rm \mathop{max}\limits_{\emph{i}\in \Theta}}$\,$\omega_{u,i}^k$.\vspace{1mm}
		\par ~~$L=\{i\in \Theta:(\widehat{m}_{{u,i},[k]}^k-\widehat{m}_{{u,j},[k]}^k)^\top(\widehat P_{{u,j},[k,k]}^k)^{-1}(\widehat{m}_{{u,i},[k]}^k-\widehat{m}_{{u,j},[k]}^k)\le\mathit{\Gamma_a}\}$.\vspace{1mm}
		\par~~$\bar\omega_\ell^k=\sum\nolimits_{i \in L}\omega_{u,i}^k$,\vspace{1mm}
		\par~~$\bar\Phi_\ell^k=\Phi_{u,j}^k$ with weight $\bar\omega_\ell^k$.\vspace{1mm}
		\par~~$\Theta=\Theta\backslash L$.\vspace{1mm}\\
		\textbf{If} $\Theta=\emptyset$, \textbf{break}\\
		if $\ell>J_{max}$ then replace $\bar\Phi_\ell^k$ by the $J_{max}$ components with largest weights.\\	
		\textbf{Output}:$\{\bar\Phi_j^k\}_{j=1}^{\text{min}\{\ell,J_{max}\}}$.\\
		\bottomrule
	\end{tabular}
\end{table}
The pruning step deletes the components with low weights and absorption step retains the component with higher weight for two closely spaced GGIW/Gaussian components. The process is given in the Table \ref{Algorithm}. 
\par To limit the increasing computational cost of handling trajectories with increasing lengths, the $L$-scan approximation \cite{Angel2019TPHD} is applied, which only updates the density of the last $L$ time and leaves the rest unaltered. It is given by approximating the covariance matrices as:
	\begin{align}
		\widehat{P}^k_{u,j}\approx\text{diag}(\widehat{P}^k_{u,j,[t_j,t_j]},\widehat{P}^k_{u,j,[t_j+1,t_j+1]},...,\widehat{P}^k_{u,j,L})
	\end{align} 
	with $u = e, p$ and $L=[k-L+1:k,k-L+1:k]$. $\widehat{P}^k_{u,j,L}\in\mathbb{R}^{Ln_x\times Ln_x}$ denotes the joint
	covariance of the $L$ last time instants, which enables the update of Gaussian densities within the $L$-scan window.
 \par  {\color{black} Besides, we also merge the GGIW components $\lambda^{S,e}_{k|k-1}\left( X \right)$,~$\lambda^{S,p,e}_{k|k-1}(X)$ \cite{Angel2021EXPMBM} and the Gaussian components $\lambda^{S,p}_{k|k-1}\left( X \right)$,~$\lambda^{S,e,p}_{k|k-1}(X)$ \cite{Vo2006PHD} in \eqref{four_term_pre} after the prediction step respectively.} %In this paper, the implementation of the PHD filter for coexisting point and extended targets can be obtained by taking the last time step estimates of the general TPHD filters, i.e., the value of $L$-scan is taken as 1\cite{Angel2019TPHD}.
\section{Numerical Study}
In this section, we will first show the tracking performance of our proposed method in a scenario in which the target type does not change in time. Then, we will focus on the performance of the proposed method considering the switching between extended and point targets. {\color{black}The metric used to evaluate the performance is the trajectory GOSPA (T-GOSPA) metric, also called trajectory metric (TM)\cite{Angel2020TM}, which is a metric for sets of trajectories that penalizes localization errors for properly detected targets, the number of missed targets, the number of false targets, and the number of track switches.}
\par We compare our proposed method with the following algorithms:
\begin{itemize}
    \item G-TPHD implemented with an $L$-scan window. It is worth noting that TPHD filter \cite{Angel2019TPHD} enables the estimation of past states of the trajectories, but a TPHD filter has the same information regarding the current state of the trajectories as a PHD filter.
    \item P-TPHD \cite{Angel2019TPHD} is the standard point-target TPHD filter.
    \item E-TPHD is the TPHD filter only considering the measurement model for extended target tracking, which is implemented by the GGIW model \cite{ExTPMB, GransXradar2015}.
    \item Tagged-PHD is a PHD filter on the set of current targets in which each component has a tag. It works similarly to the G-TPHD filter implementation but only keeping information on the set of current targets. It then links estimation with the same tag at different time steps to generate trajectories, similarly to \cite{Panta-2009-taggedPHD}.

\end{itemize}
\subsection{Simulation Scenario I}
\par The first simulation aims at showing a brief traffic environment with the size of $[ {{\rm{ - 150}},{\rm{150}}}]\text{m}\times[ {{\rm{ -150}},{\rm{150}}} ]\text{m}$, where three extended targets and four point targets are moving for 100 seconds (Table \ref{Target States}). The target state {\color{black}vector} is given as $x = {[ {{p_x},{p_y},{{\dot p}_x},{{\dot p}_y}} ]^\top}$ including the position (with unit: $m$) and velocity information (with unit: $m/s$). {\color{black}The ground truth trajectories are generated according to a nearly constant velocity motion model by using $F$ and $Q$ below. The initial positions of these targets as well as their type and presence times are given in Table II.} The observation {\color{black}vector} $z = {[ {{z_x},{z_y}}]^\top}$ includes the position information. The single target transition model is given as
\begin{align*}
	F =& \left[ {\begin{array}{*{20}{c}}
			{{I_2}}&{{I_2}\delta t}\\
			{{0_2}}&{{I_2}}
	\end{array}} \right] \qquad	Q = \sigma _v^2\left[ {\begin{array}{*{20}{c}}
			{\frac{{\delta {t^4}}}{4}{I_2}}&{\frac{{\delta {t^3}}}{2}{I_2}}\\
			{\frac{{\delta {t^3}}}{2}{I_2}}&{\delta {t^2}{I_2}}
	\end{array}} \right]\\
	H =& \left[ {\begin{array}{*{20}{c}}
			{{I_2}}&{{0_2}}
	\end{array}} \right]\qquad~~~	R = \sigma _\varepsilon ^2{I_2}
\end{align*}
where ${I_2}$ represents the $2\times 2$ unit matrix, {\color{black}$0_2$ represents the $2\times 2$ zero matrix, $\sigma_v = 0.01\text{m/s}^{2}$, $\sigma _\varepsilon = 1 \text{m}^2$.} The notation $\delta t = 1\text{s}$ denotes the sampling period. 
\begin{table}[!t]
	\centering
	\caption{The Simulated Target States}
	\label{Target States}	
	\tiny
	\begin{tabular}{c|c|c|c|c}
		\hline
		\hline
		&Kind&Initial State&Birth Time$/s$&Death Time$/s$\\
		\hline
		Target 1&Extended&$\left[-85,-7.5,2,0\right]^{\top}$&1&100\\	 \hline
		Target 2&Point&$\left[65,5,-2.1,0\right]^{\top}$&7&100\\	 \hline
		Target 3&Point&$\left[9.5,-55,0,2.4\right]^{\top}$&10&80\\
		\hline
		Target 4&Extended&$\left[-5,110,0,-2.5\right]^{\top}$&20&70\\	 \hline
		Target 5&Point&$\left[-10,90,0,-2.8\right]^{\top}$&25&100\\  \hline
		Target 6&Extended&$\left[-6,-51,0,-1.2\right]^{\top}$&55&100\\  \hline
		Target 7&Point&$\left[55,10,-0.65,-0.35\right]^{\top}$&60&100\\  \hline
	\end{tabular}
\end{table}
\begin{table}[!t]
	\centering
	\caption{Filter Parameters}
	\label{Filter Parameters}
	\scriptsize
	\begin{tabular}{|c c|c|}
		\hline
		\multicolumn{2}{|c|} {G-TPHD filter}&Value\\ \hline
		Probability of survival &$p^S$ &0.99\\	 
		Probability of detection &$p^D$ &0.95\\	
		Mean number of clutter/per scan &$\lambda_c$  &5\\
		Threshold of pruning &$\mathit{\Gamma_p}$ &$10^{-5}$\\	
		Threshold of absorption & $\mathit{\Gamma_a}$ & $4$\\  
		Maximum PHD components & $J_{max}$ & $100$\\  
		Value of $L$-scan & $L$ & $5$\\ 
		Dimension of $\mathbb{S}_+^{d}$& $d$ & 2\\
		Birth weight of target& $[\omega_{\gamma^e},\omega_{\gamma^p}]$ & $[0.05,0.05]$\\  
		Birth parameters of Gamma & $a,b$ & $8,1$\\  
		Birth parameters of Inverse Wishart & $v,V$ & $10,I_d$\\  	
		Measurement rate (Eq. \eqref{mu})& $\mu$ & $1.05$\\
		Correlation constant (Eq. \eqref{tau})& $\tau$ & $5.48$\\
		Transformation matrix (Eq. \eqref{M}) & $M(\cdot)$ & $I_d$\\
		Parameters of DBSCAN & $\tau_{db},\Gamma_{min},\Gamma_{max}$ & $0.1,0.1,5$\\ 
		\hline 
	\end{tabular}
\end{table}
\par
\begin{figure}[!t]
	\centering
	\includegraphics[width=2.8in]{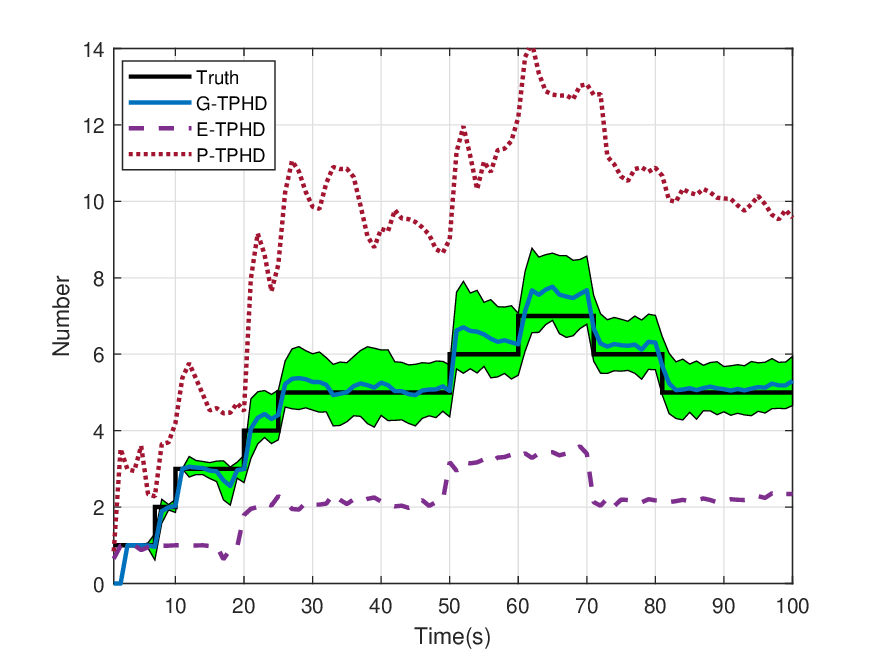}
	\caption{ This figure shows the estimated number of alive trajectories for the G-TPHD ($L=5$), E-TPHD and P-TPHD filters. The green area represents the fluctuation of the G-TPHD estimator within one standard deviation. It is worth noting that the Tagged-PHD filter estimates the same number of alive trajectories with the G-TPHD filter.}
	\label{number}
\end{figure}

\begin{figure}[!t]
	\centering
	\includegraphics[width=3.0in]{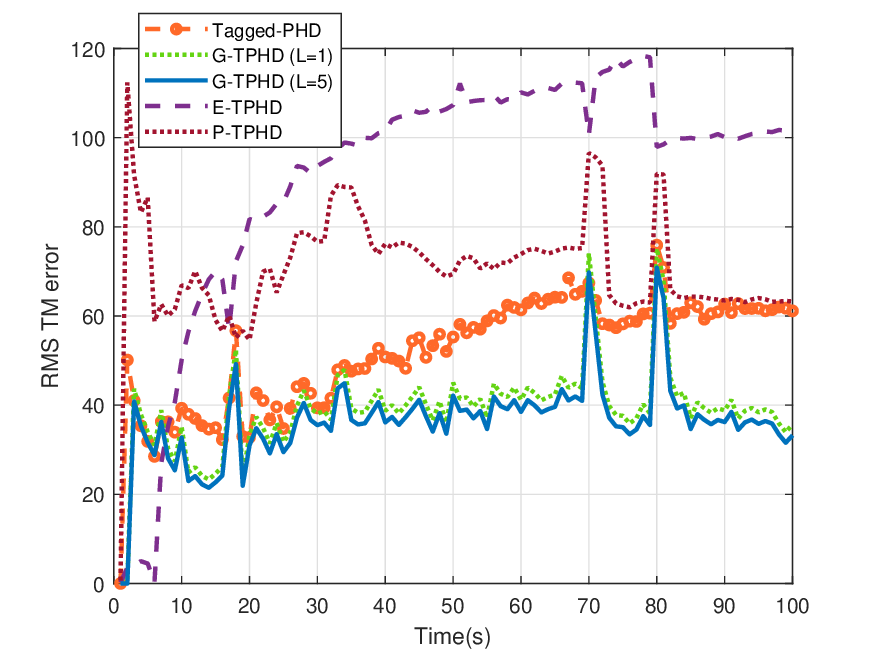}
	\caption{{\color{black}The figure shows the RMS TM error for the Tagged-PHD, G-TPHD ($L=1$ \text{and} $L=5$), E-TPHD and P-TPHD filters}}
	\label{TM}
\end{figure}

\begin{figure}[!t]
	\centering
	\includegraphics[width=3.0in]{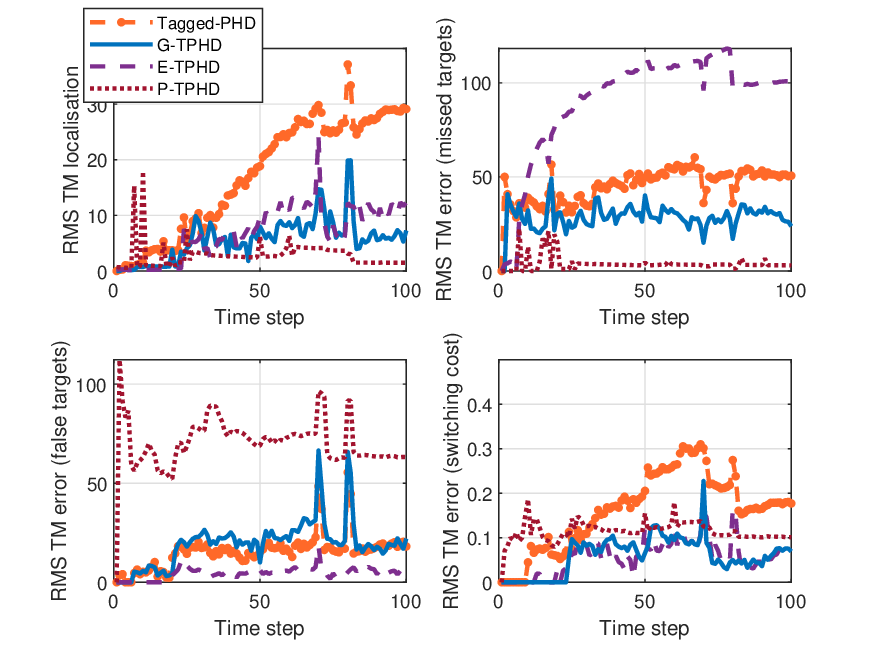}
	\caption{The figure shows the decomposition of RMS TM error (including localization, missed targets, false targets and switching error) for the Tagged-PHD, G-TPHD ($L=5$), E-TPHD and P-TPHD filters.}
	\label{TM_De}
\end{figure}

 The parameters in the proposed G-TPHD filter are given in Table \ref{Filter Parameters}. {\color{black}In a practical setting, the parameters of the measurement model can be obtained by iteratively maximising the likelihood \cite{Angel2024TPMBM_traffic}}. For measurement clustering, we generate possible partitions $\mathcal{P}$ of measurement set $\mathbf{z}$ by using the DBSCAN algorithm \cite{DBSCAN} with distance thresholds between $\Gamma_{min}$ and $\Gamma_{max}$, with a step size of $\tau_{db}$ (unit: \text{m}). The minimum number of points to form a region
is set to 1 to capture point target measurements. Meanwhile, we only keep the unique partitions among all possible generated partitions and we obtain the unique subsets of measurements in these partitions. Besides, the number of clutter measurements per scan is Poisson distributed and the clutter location is concentrated within the road region. The parameters of the root mean square (RMS) TM error are $p=2,c=100,\gamma=1$. The basic matrix is the Gaussian Wasserstein matrix \cite{Yang2016TM}. All estimates in Figs. \ref{number} -- \ref{TM_De} are averaged by 300 Monte Carlo runs.
\begin{figure}[!t]
	\centering
	\includegraphics[width=2.8in]{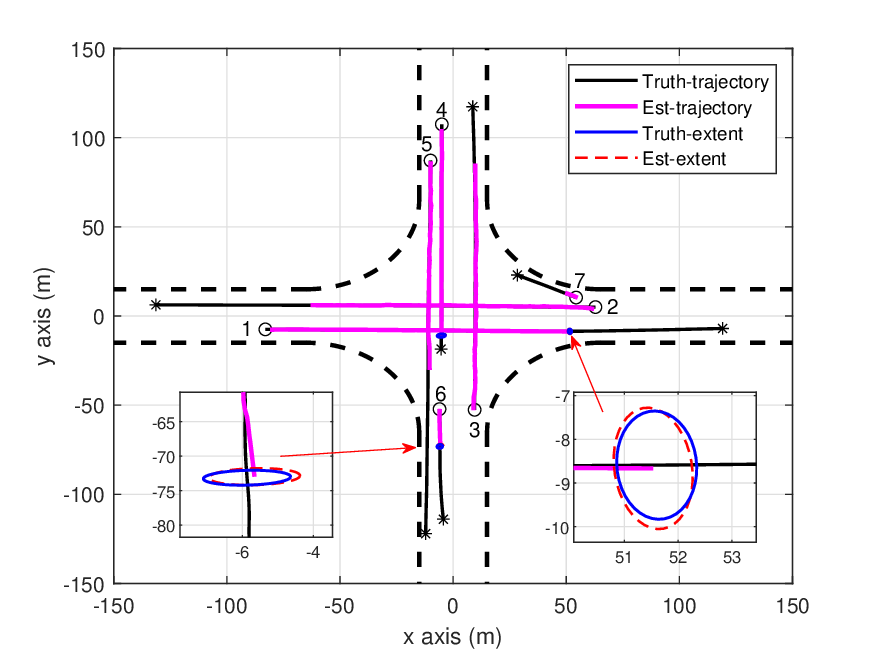}
	\caption{This figure shows a simulated traffic environment, where seven different targets are moving within 100s. The radar sensor is set at the position $(0,0)$. The start and end points for each true trajectory are marked by $\text{o}$ and $*$, respectively. The black lines denote the true trajectories for all targets during the whole time period (100s). Using the G-TPHD filter, we present the estimated target trajectories (purple lines) and target extent (red dashed lines) at the time step 67s. The true extent model for extended targets is marked through blue lines.}
	\label{state}
\end{figure}
%	\multicolumn{5}{|c|}{The RMS TM Error and its Decomposition}\\  

\par It can be seen from Fig. \ref{state} that the G-TPHD filter can track point and extended targets at the same time, and output their current position and trajectory information. From Table \ref{TM_error_table} and Figs. \ref{number} -- \ref{TM_De}, the G-TPHD filter is more accurate in estimating the target trajectory and the number of targets at each time step compared to the E-TPHD, P-TPHD and Tagged-PHD filters. For the advantages of the proposed G-TPHD ($L=5$) filter over the G-TPHD ($L=1$) version in Fig. \ref{TM}, the former considers using current measurements to update a larger length of trajectory so that a better estimate can be obtained. {\color{black}The average runtime for the G-TPHD filter with $L =1$ is 3.5s, and with $L =5$ is 4.2s. There is a trade-off between computational burden and performance (higher $L$ implies higher performance with higher computational burden). How much benefit one gets from increasing the $L$-scan window size depends on the specific dynamic and measurement models.}
\par The reason why the proposed G-TPHD filter is better than the other three counterparts is explained by looking at Fig. \ref{TM_De}, which shows the decomposition of RMS TM error. For the P-TPHD filter that only considers point target tracking, extended targets with multiple measurements will cause more false target estimates. On the other hand, for the E-TPHD filter, the hypothesis of point target has a smaller posterior probability, resulting in more missed target errors. For the proposed G-TPHD filter, the performance is more robust because it considers the space of coexisting point and extended targets, and a general measurement likelihood function. {\color{black}We can also see that the Tagged-PHD filter has a higher cost for localization, missed targets and track switches than the proposed G-TPHD filter. This is because because it is not a principled Bayesian method to estimate sets of trajectories. For instance, each Gaussian PHD component represents a set of targets distributed as a PPP (not a single target), so there can be multiple potential targets with the same tag.}
\begin{table}[!t]
	\centering
	\caption{The RMS TM Error and Its Decomposition (Normalized by Time Window) }
	\label{TM_error_table}
	\scriptsize
	\begin{tabular}{|c|c|c|c|c|c|} 
		\hline
		\hline
		&Mean&Localization&Miss&False&Switch\\ \hline
		Tagged-PHD&52.25&17.23&45.30&15.52&0.163\\   \hline
		G-TPHD (L=1)&37.75&6.78&30.31&20.56&0.045\\   \hline
		G-TPHD (L=5)&35.19&5.64&27.45&18.44&0.031\\ \hline
		E-TPHD&91.16&7.01&90.67&4.70&0.059\\ \hline
		P-TPHD&71.20&2.89&3.89&70.53&0.110\\ \hline
	\end{tabular}
\end{table}
\par {\color{black}Besides, we also test the performance (RMS TM error) of the G-TPHD filter by changing the probability of detection. It is shown in Table. \ref{pd_table}, where the we have three different values of detection probability (True $p^D$) used to generate the measurement and three different values of detection probability (model $p^D$) used in the filter. We can see that, the filter works well if there is not a big gap ($\sim$0.1) between model $p^D$ and true $p^D$. For lower $p^D$, performance goes down (even if the model is perfectly matched).}
\begin{table}[!t]
		\centering
	\caption{The RMS TM Error of the G-TPHD filter under Different Detection Probability}
	\label{pd_table}
	\scriptsize
        \begin{tabular}{|c|c|c|c|}
                \hline
                \multirow{2}{*}{\begin{tabular}{c} \rotatebox{45}{\textbf{}} \\ \textbf{Model $p^D$} \end{tabular}} & \multicolumn{3}{c|}{\textbf{True $p^D$}} \\
                \cline{2-4}
                & 0.9 & 0.8 & 0.7\\
                \hline
                0.95 & 43.80 & 59.10 & 74.12 \\
                \hline
                0.9 & 40.96 & 49.37 & 62.66 \\
                \hline
                0.8 & 42.73 & 45.34 & 52.89 \\
                \hline
        \end{tabular}
\end{table}
\subsection{Simulation Scenario II}

\begin{figure}[!t]
    \centering
    \begin{subfigure}[b]{0.22\textwidth}
        \centering
        \includegraphics[width=\textwidth]{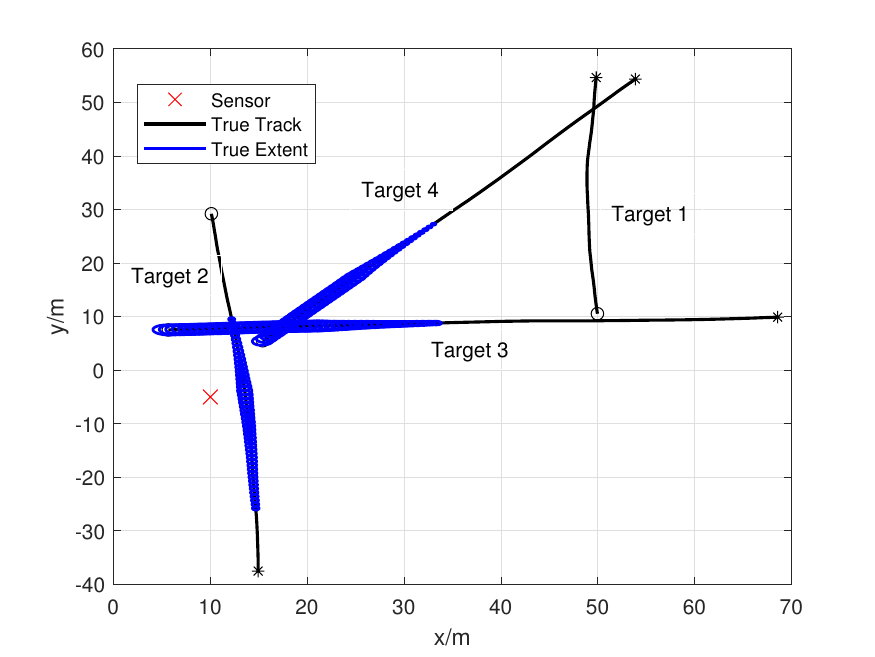}
        \caption{Truth}
        \label{fig:img1}
    \end{subfigure}
    \hspace{0.1cm}
    \begin{subfigure}[b]{0.22\textwidth}
        \centering
        \includegraphics[width=\textwidth]{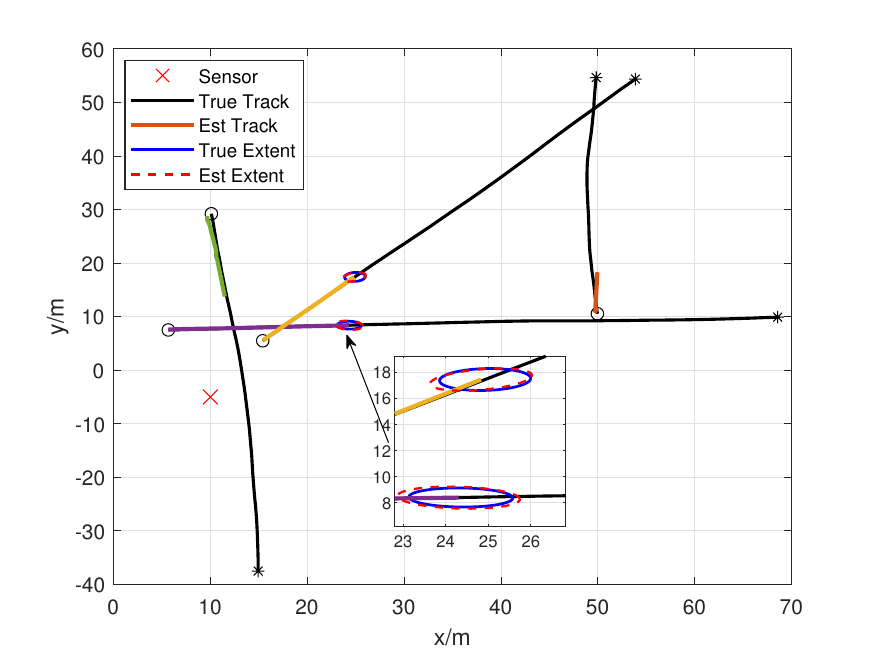}
        \caption{Time Step 30}
        \label{fig:img2}
    \end{subfigure}
    \\[-0.2cm]
    \begin{subfigure}[b]{0.22\textwidth}
        \centering
        \includegraphics[width=\textwidth]{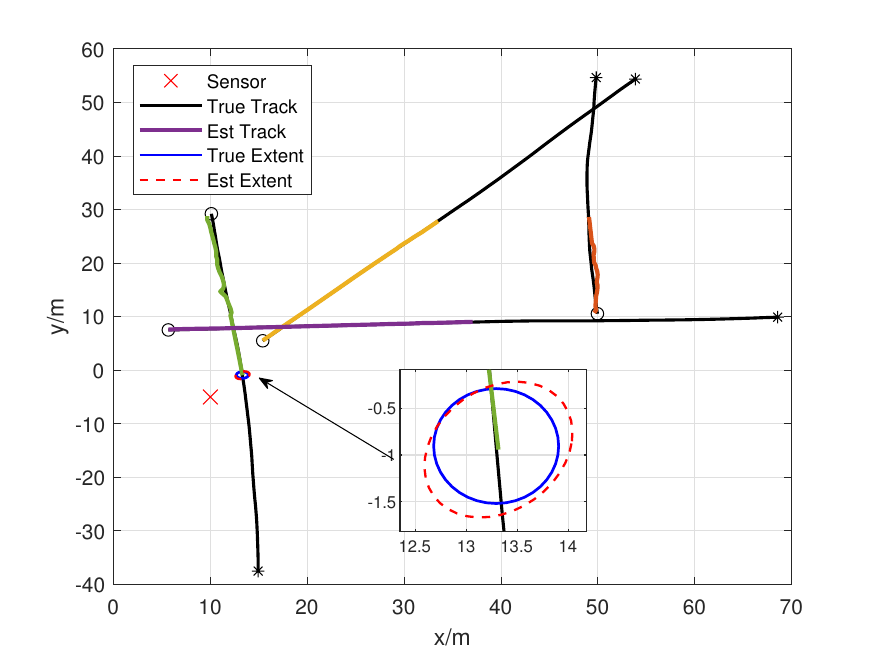}
        \caption{Time Step 50}
        \label{fig:img3}
    \end{subfigure}
    \hspace{0.1cm}
    \begin{subfigure}[b]{0.22\textwidth}
        \centering
        \includegraphics[width=\textwidth]{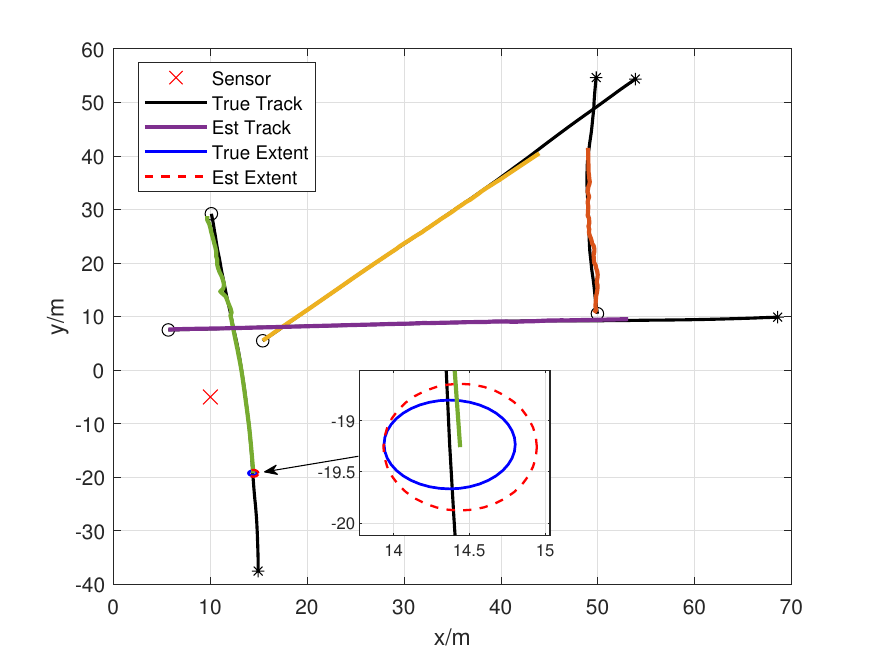}
        \caption{Time Step 75}
        \label{fig:img4}
    \end{subfigure}
    \caption{(a) shows the ground truth of targets' trajectories, extent and how their type changes. (b), (c), (d) show the estimates of targets' trajectories and extent at different time steps.}
    \label{fig:four_images}
\end{figure}
\begin{figure}[!t]
    \centering
    \begin{subfigure}[b]{0.22\textwidth}
        \centering
        \includegraphics[width=\textwidth]{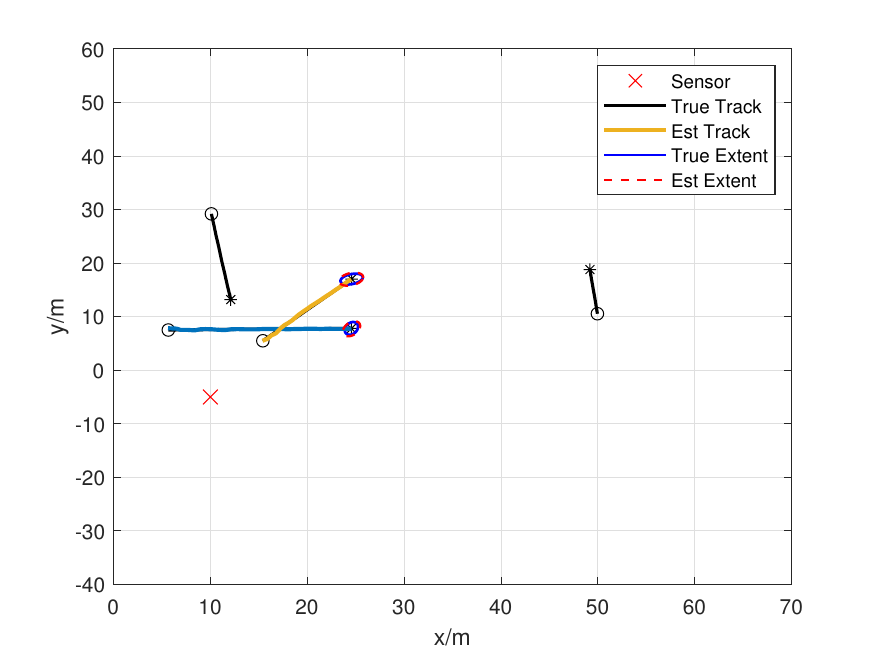}
        \caption{E-TPHD estimation at the time step 30s. It only successfully tracks two extended targets and loses other two point targets.}
        \label{E1}
    \end{subfigure}
    \hspace{0.1cm}
    \begin{subfigure}[b]{0.22\textwidth}
        \centering
        \includegraphics[width=\textwidth]{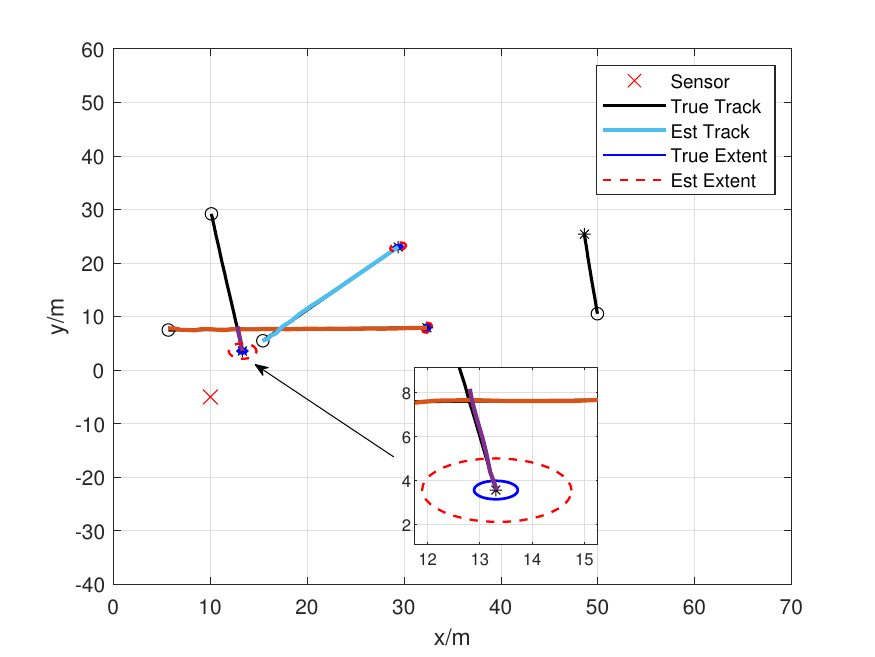}
        \caption{E-TPHD estimation at the time step 42s. It begins to track the purple target which is transferred from a point target.}
        \label{E2}
    \end{subfigure}
    \caption{Tracking results of E-TPHD at different frames in one sample.}
    \label{two_images_Extended}
\end{figure}
\begin{figure}[!t]
	\centering
	\includegraphics[width=3.0in]{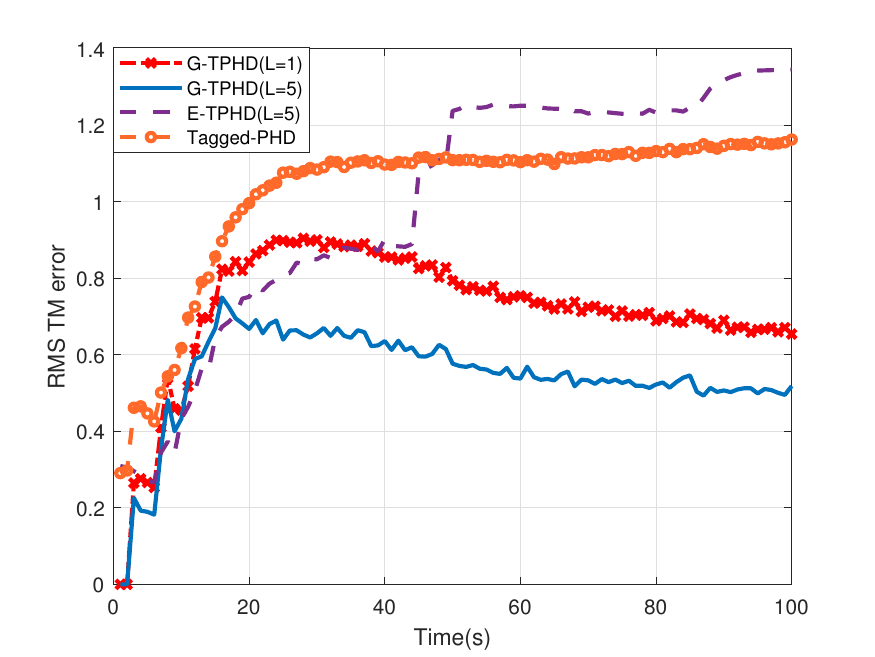}
	\caption{The RMS TM error for the Tagged-PHD, G-TPHD ($L=1,5$) and E-TPHD ($L=5$).}
	\label{TM2}
\end{figure}
\begin{figure}[!t]
    \centering
    \begin{subfigure}[b]{0.22\textwidth}
        \centering
        \includegraphics[width=\textwidth]{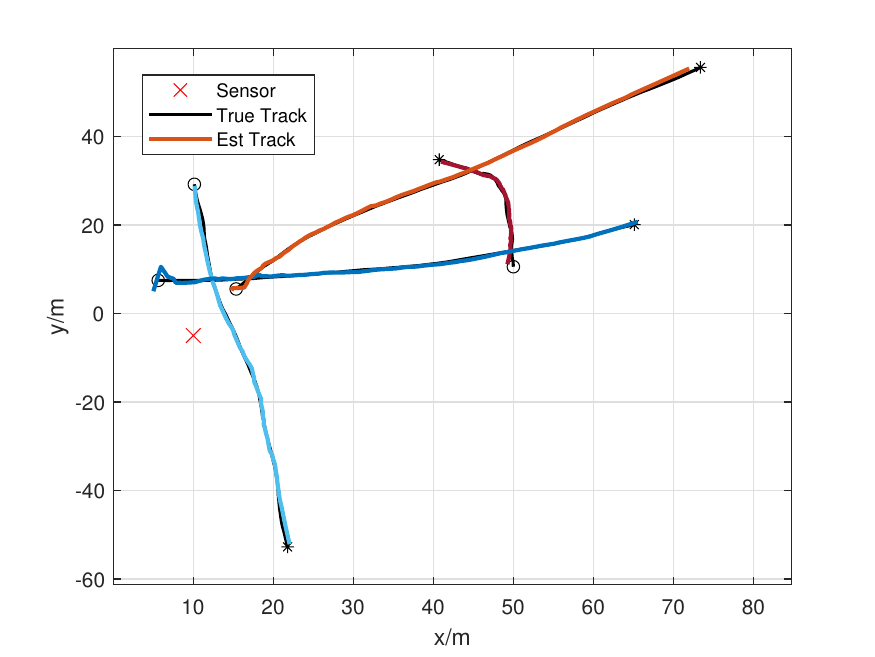}
        \caption{$\sigma_v = 0.05$, RMS trajectory localization error = 0.48.}
        \label{fig:sigma0.05}
    \end{subfigure}
    \hspace{0.1cm}
    \begin{subfigure}[b]{0.22\textwidth}
        \centering
        \includegraphics[width=\textwidth]{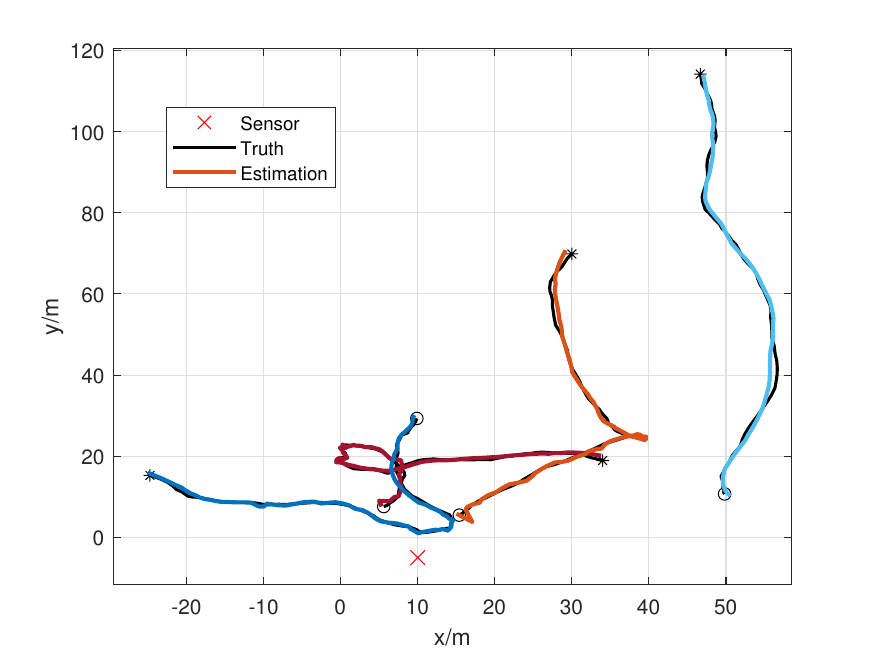}
        \caption{$\sigma_v = 0.3$, RMS trajectory localization error = 0.61}
        \label{fig:sigma0.3}
    \end{subfigure}
    \caption{Tracking results under different motion noise levels. The localization error is averaged by 300 runs.}
    \label{fig:two_images}
\end{figure}

{\color{black}In this subsection, we focus on the performance when there exists switching between point and extended targets. There are four targets (Fig. \ref{fig:four_images}), where 
\begin{itemize}
    \item ``Target 1" is always point target.
    \item ``Target 2" is a point target first (10s-35s) and then changes into an extended target (35s - 85s) and then switches back ($>$85s).
    \item ``Target 3" is an extended target first (7s-50s), and then changes into a point target ($>$50s).
    \item ``Target 4" is an extended target (1s-45s), and then changes into a point target ($>$45s). 
\end{itemize}  
The ground truths of these four trajectories are generated according to a nearly constant velocity motion model (See Fig.~\ref{fig:img1}). The clutter rate is set to $\lambda_c = 5$ per frame, and the motion noise is set as $\sigma_v = 0.02\text{m/s}^2$, and the measurement noise is given as $\sigma_\epsilon = 1\text{m}^2$. All filter parameters are also the same as those in Simulation I. The parameters of the RMS TM error is set as $p=2, c=1, \gamma=1$. For the transition probabilities $s_e$ and $s_p$, we set both $s_e$ and $s_p$ to 0.9.
\par In the G-TPHD filter, the point and extended target trajectories are described by different components $\lambda^k_{p}$ and $\lambda^k_e$, where each component only belongs to one type of target. The filter does not provide a probability of classification, instead, it directly provides the expected number of point and extended targets in a given area.

\par Our design performs well in the scenario (Figs.\ref{fig:four_images}) where the target is more likely to be an extended target if it is closer to the sensor, and to be a point target if it is farther away. This is not solely due to the transition density in the prediction step but also due to the general measurement model. Specifically, the transition density $g_e(\cdot)$ models the switching between these two types as well as generating corresponding hypothesis. The general measurement model $L_{\mathbf{z}_{k}}(\cdot)$ helps us to judge which hypothesis has bigger probability.
% \par From Figs.\ref{fig:four_images} and \ref{class}, we can see that the proposed method can output correct trajectory estimation and classification. Although, there exists fluctuations at the beginning for classification, but the output generally keeps stable with tracking time increasing.
\par From Fig.~\ref{TM2}, we obtain similar results as simulation I. With increasing number of point targets (roughly after 45s), E-TPHD becomes worse and worse as there is a measurement model mismatch between point and extended target. The extended target measurement model lacks the support for point target. Because point targets can at most generate one measurement while extended targets can generate multiple measurements, modeled by a Poisson point process. It can be seen from Fig.~ \ref{two_images_Extended} that, the E-TPHD filter fails to track the point targets at 30s and only when ``Target 2'' changes into extended target at 42s, E-TPHD gradually recovers the ability to track it. This fact illustrates the importance to design a general measurement likelihood. Meanwhile, we can also see that the G-TPHD filter generally performs better than Tagged-PHD filter and this gap will become larger with increasing $L$. Because the G-TPHD filter will update the latest trajectory with length $L$. Meanwhile, we can also see from Fig.\ref{fig:two_images} that the filter can still output robust trajectory state estimation under high motion noise.}

\subsection{Experimental Scenario}
\begin{figure}[!t]
	\centering
	\includegraphics[width=3.0in]{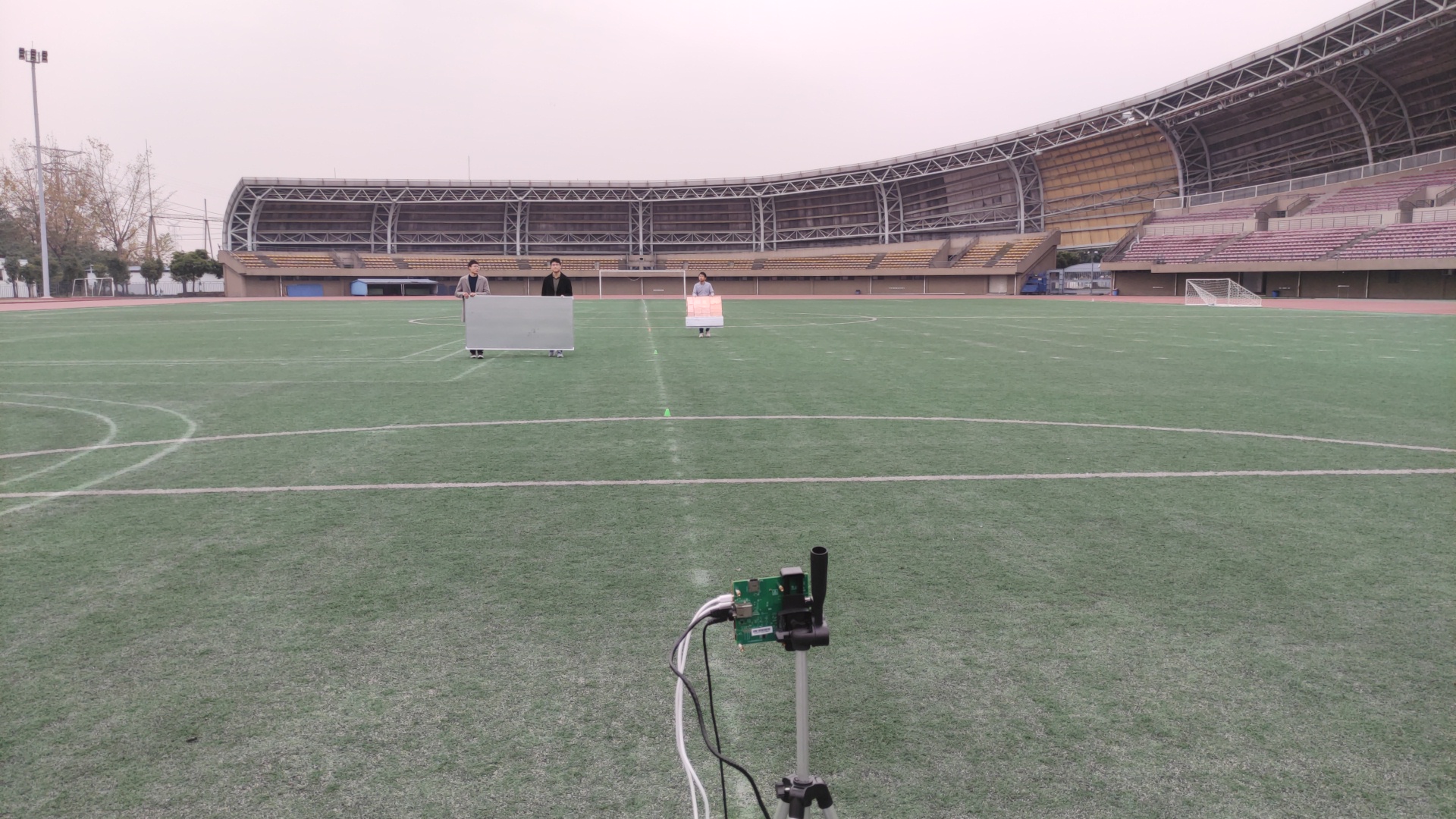}
	\caption{ The experimental scenario.}
	\label{Sc}
\end{figure}
\begin{figure}[!t]
	\centering
	\includegraphics[width=3.0in]{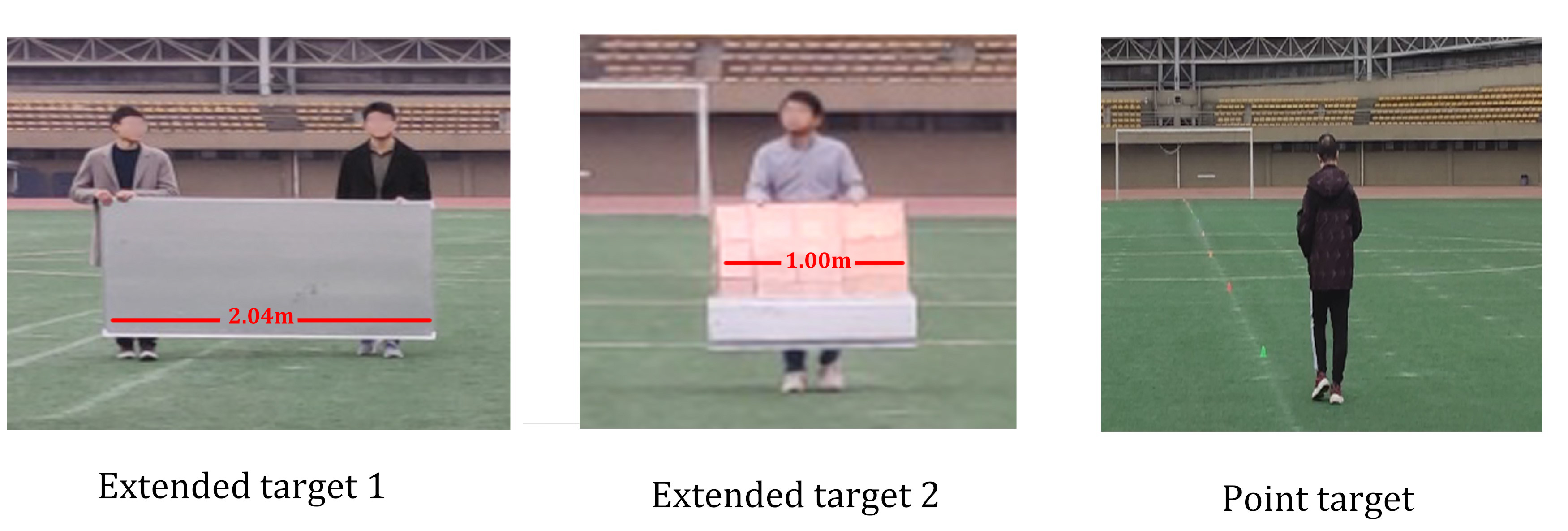}
	\caption{ The targets in the tracking scenario (two kinds of extended target with different size and two point targets).}
	\label{target}
\end{figure}
\begin{figure}[!t]
	\centering
	\includegraphics[width=3.2in]{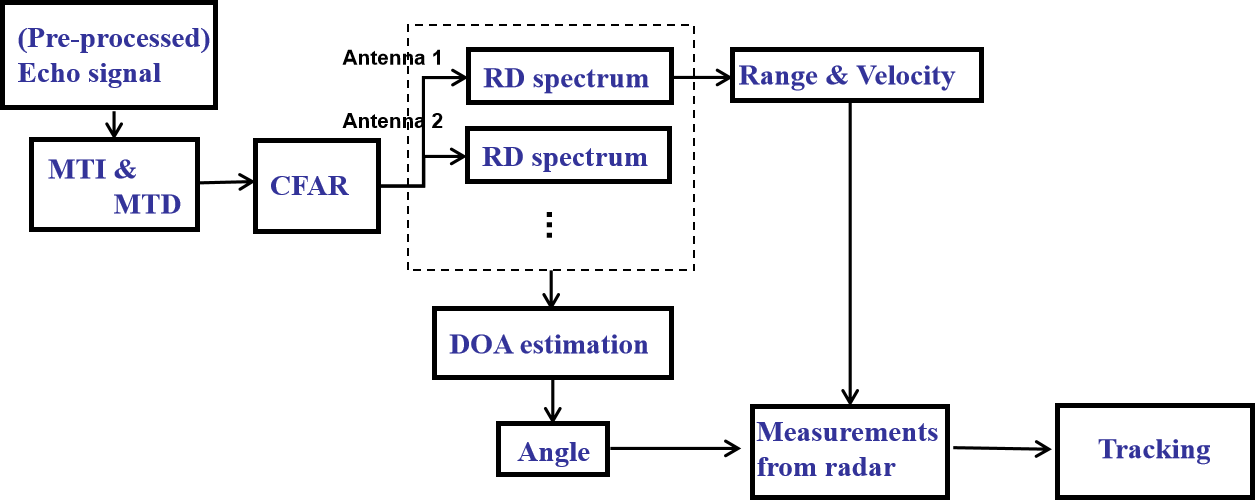}
	\caption{The signal processing flow for collected data from Fig \ref{Sc}.}
	\label{flow}
\end{figure}
In this subsection, we present the signal processing and tracking results from the measured radar data. The experimental scenario is from the playground of the university campus, as shown in Fig. \ref{Sc}. There are two extended targets and two point targets. Both of them move in an approximately straight line. The extended targets consist of people and a metal board, while the point target is a single person, as shown in Fig. \ref{target}. The transmitting signal is frequency modulated continuous wave (FMCW) with the start frequency 77GHz. The Bandwidth is set as 800MHz. The number of chirp loops and ADC samples are 128 and 256, respectively. The structure of the radar equipment consists of 1 transmitting antenna and 4 receiving antennas. The total length of the data used in tracking is 7s (70 frames) and the filter parameters are the same with those in Subsection A.

\par
\begin{figure}[!t]
	\centering
	\includegraphics[width=3.4in]{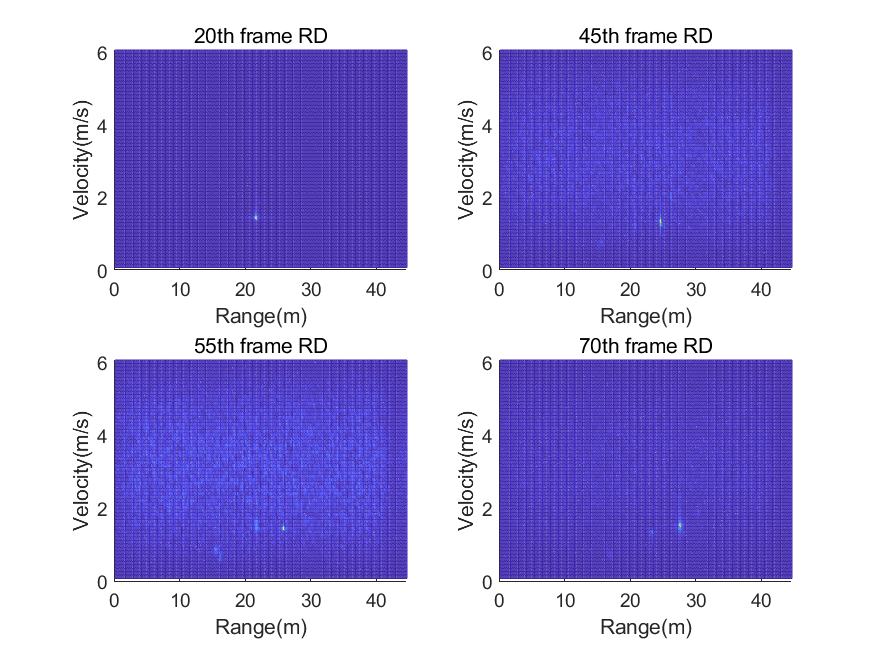}
	\caption{ The figure shows the range and Doppler (RD) spectrum of one radar channel at the 20th, 45th, 55th and 70th frames. A brighter color represents a stronger signal.}
	\label{RD}
\end{figure}

\begin{figure}[!t]
	\centering
	\includegraphics[width=3.4in]{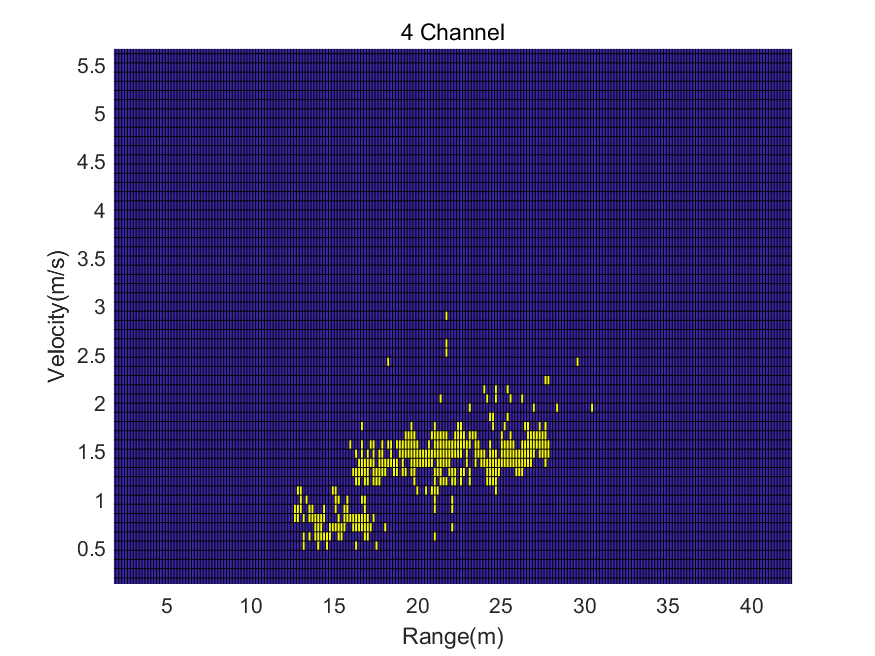}
	\caption{ It shows the result after CFAR, which sums over 4 channels of radar RD spectrum during the whole tracking period (70 frames). Here, the amplitude of all selected peaks after CFAR is set as 1.}
	\label{CFAR}
\end{figure}

\begin{figure}[!t]
	\centering
	\includegraphics[width=3.0in]{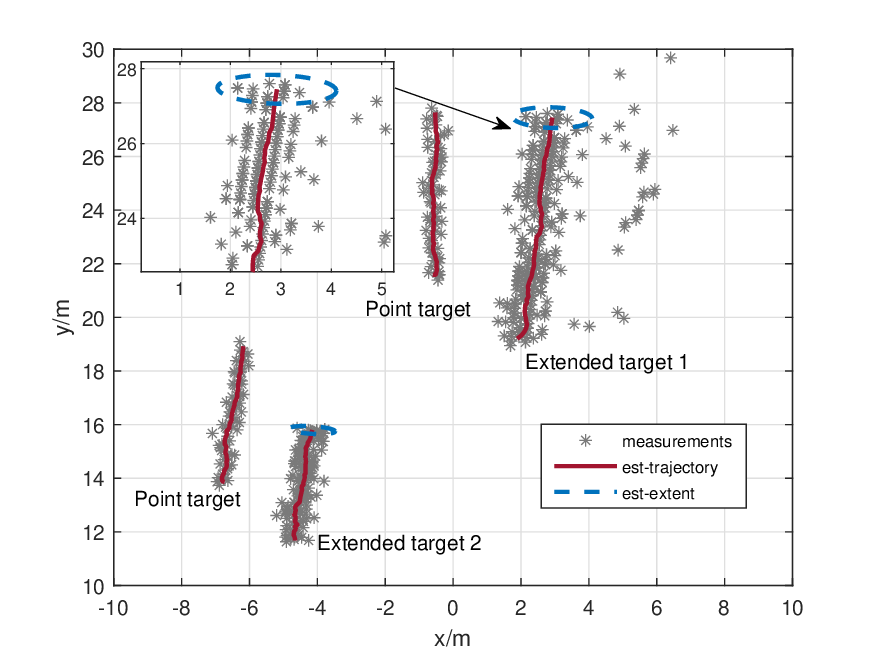}
	\caption{ Two point and two extended targets are moving away from the radar at the origin during the time period from the 1st frame to the 70th frame. The tracking results are represented with red lines and the estimates of the target extent are represented with blue dotted lines. Asterisks represent measurements.}
	\label{est}
\end{figure}

 The signal processing steps are given by the flow chart Fig. \ref{flow}. The Moving Target Indicator (MTI) filter is achieved by the transfer function $H(z)=1-z^{-1}$ to eliminate the static target and clutter. The Doppler and range information is obtained by doing fast Fourier transform (FFT) for the slow and fast time dimension respectively. After obtaining the range and Doppler (RD) spectrum, we use the constant-false-alarm-rate (CFAR) technique to extract peaks by setting the threshold. Here, we adopted a square and two-dimensional Cell-Averaging-CFAR (CA-CFAR) \cite{CACFAR}. The constant false alarm rate is set as $P_{fa}=1 \times10 ^{-6}$, the number of protection units is $N_p=48$, and the number of reference units is $N_r=312$. The RD spectrum and its result after CFAR are given in Figs. \ref{RD} and \ref{CFAR}. To get the angular information, we do FFT directly on the angular dimension. Then by combining the distance information after CFAR and angular information, we can obtain the X-Y position information of targets and use it as the measurement set for the proposed filter to track. The estimated trajectory result is shown in Fig. \ref{est}. It can be seen that the proposed filter can distinguish point target and extended target well. Besides, based on the ellipse model for target extent, we can make the average estimates for the widths of two extended targets, which are shown in Table \ref{Width}. 
\begin{table}[!t]
	\centering
	\caption{Estimated Width for the Extended Targets}
	\label{Width}
	\scriptsize
	\begin{tabular}{|c|c|c|}
		\hline
		\hline
		Category&True Width (m)&Estimated Width (m)\\
		\hline
		Extended Target 1&2.04&2.24\\	 \hline
		Extended Target 2&1.00&1.25\\	 \hline
	\end{tabular}
\end{table}

\section{Conclusion}
In this paper, we have derived a TPHD filter for general target-generated measurements by approximating the posterior density by a Poisson density using direct KLD minimization without using PGFLs. The proposed TPHD filter is able to output the trajectory estimates of both point and extended targets simultaneously. In addition, we have proposed an implementation of this filter based on a Gamma Gaussian Inverse Wishart mixture model. Considering the tracking efficiency, a $L$-scan approximate version is proposed to reduce the computational cost. Finally, the simulation and experimental results proved that the proposed TPHD filter has a robust performance in coexisting point and extended target scenarios. {\color{black}A line of future work is to develop and implement TPHD filters for more general dynamic models, including the case in which the transition density depends not only on the current state, but in past states of the trajectory.}
\appendices

\section{\label{sec:AppendixA}}
We aim to prove Proposition 1 by finding the Poisson density that best fits the posterior by minimizing the KLD, similar to the approaches in
\cite{Angel2019TPHD,Angel2015KLD}. We commence by writing the predicted PHD for targets as
\begin{align}\label{target_PHD}
	\lambda_{k|k-1}\left(x\right) & =\overline{\lambda}_{k|k-1}\breve{\lambda}_{k|k-1}\left(x\right),
\end{align}
where
\begin{align}
	\overline{\lambda}_{k|k-1} & =\int\lambda_{k|k-1}\left(x\right)dx,\\
	\breve{\lambda}_{k|k-1}\left(x\right) & =\frac{\lambda_{k|k-1}\left(x\right)}{\overline{\lambda}_{k|k-1}}.
\end{align}
The notation $\overline{\lambda}$ indicates the expected number of targets. Considering the target trajectory, its corresponding PHD $\lambda_{k|k-1}\left(X\right)$ possesses the similar decomposition and the same expected number as the targets (as the PHD only considers alive trajectories). That is, 
\begin{align}
	\int\lambda_{k|k-1}\left(x\right)dx=\int\lambda_{k|k-1}\left(X\right)dX.
\end{align}
Given the PHD $\lambda_{k|k-1}\left(X\right)$, the corresponding
PPP density is
\begin{align}
	f_{k|k-1}\left(\left\{ X_{1},...,X_{n}\right\} \right) & =e^{-\overline{\lambda}_{k|k-1}}\overline{\lambda}_{k|k-1}^{n}\prod_{j=1}^{n}\breve{\lambda}_{k|k-1}\left(X_{j}\right),\label{eq:PPP_prior}
\end{align}
which can be also applied in the PHD of targets $f_{k|k-1}\left(\left\{ x_{1},...,x_{n}\right\} \right)$ by only changing the notation $X$ to $x$.

\subsection{Preliminary lemmas}

The proof of the general TPHD filter update makes use of the following
lemmas.
\begin{lem}
	\label{lem:Equality_functionals1}Given a set $\mathbf{z}$, and a
	partition $Q$ of $\mathbf{z}$ and real-valued functions $\lambda\left(\cdot\right)$
	and $\tau\left(\cdot\right)$ such that
	\begin{align*}
		\lambda^{\mathbf{z}} & =\begin{cases}
			\prod_{z\in\mathbf{z}}\lambda\left(z\right) & \mathbf{z}\neq\emptyset\\
			1 & \mathbf{z}=\emptyset
		\end{cases}\\
		\tau^{Q} & =\begin{cases}
			\prod_{\mathbf{w}\in Q}\tau\left(\mathbf{w}\right) & Q\neq\emptyset\\
			1 & Q=\emptyset
		\end{cases}
	\end{align*}
	the following equality holds
	\begin{align}
		\sum_{\mathbf{y}\subseteq\mathbf{z}}\lambda^{\mathbf{z}\setminus\mathbf{y}}\sum_{\mathcal{Q}\angle\mathbf{y}}\tau^{\mathcal{Q}} & =\sum_{\mathcal{Q}\angle\mathbf{z}}\left(\kappa+\tau\right)^{\mathcal{Q}},\label{eq:equality_functionals1}
	\end{align}
	where $\mathbf{z}\setminus\mathbf{y}$ denotes the set difference
	between $\mathbf{z}$ and $\mathbf{y}$, and
	\begin{align*}
		\kappa\left(\mathbf{w}\right) & =\delta_{1}\left[|\mathbf{w}|\right]\prod_{z\in\mathbf{w}}\lambda\left(z\right).
	\end{align*}
\end{lem}
Lemma \ref{lem:Equality_functionals1} is proved in Appendix C.
\begin{lem}
	\label{lem:Equality_functionals2}Given two real-valued functions
	$f\left(\cdot\right)$ and $g\left(\cdot\right)$ defined for all
	$\mathbf{w}\subseteq\mathbf{z}_{k}:|\mathbf{w}|>0$, the following
	relation holds
	\begin{align}
		\sum_{\mathbf{w}\subseteq\mathbf{z}_{k}:|\mathbf{w}|>0}f\left(\mathbf{w}\right)\sum_{\mathcal{Q}\angle\mathbf{z}_{k}\setminus\mathbf{w}}g^{\mathcal{Q}} & =\sum_{\mathcal{P}\angle\mathbf{z}_{k}}g^{\mathcal{P}}\sum_{\mathbf{v}\in\mathcal{P}}\frac{f\left(\mathbf{v}\right)}{g\left(\mathbf{v}\right)}.\label{eq:equality_functionals2}
	\end{align}
\end{lem}
Lemma \ref{lem:Equality_functionals2} is proved in Appendix D.

\subsection{Density of the measurement}

Given the PPP prior (\ref{eq:PPP_prior}), the density of the measurement
is the union of independent clutter generated measurements and target
generated measurements, i.e., the density is target dependent when considering trajectory $f\left(\mathbf{z}|\left\{ x_{1},...,x_{n}\right\} \right)$.
\subsubsection{Target-generated measurements}
Applying the convolution formula, the density of the target-generated
measurements given the set of targets $\mathbf{x}$ is
\begin{align}
	f\left(\mathbf{z}|\left\{ x_{1},...,x_{n}\right\} \right) & =\sum_{\mathbf{z}_{1}\uplus...\uplus\mathbf{z}_{n}=\mathbf{z}}\prod_{j=1}^{n}f\left(\mathbf{z}_{j}|x_{j}\right).
\end{align}
Then, the target-generated measurements have density
\begin{align}
	l_{k|k-1}^{T}\left(\mathbf{z}\right) & =\int f\left(\mathbf{z}|\mathbf{x}\right)f_{k|k-1}\left(\mathbf{x}\right)\delta\mathbf{x}\notag\\
	& =\sum_{n=0}^{\infty}\frac{1}{n!}\int\sum_{\mathbf{z}_{1}\uplus...\uplus\mathbf{z}_{n}=\mathbf{z}}\prod_{j=1}^{n}f\left(\mathbf{z}_{j}|x_{j}\right)\notag\\
	& \quad\times e^{-\overline{\lambda}_{k|k-1}}\overline{\lambda}_{k|k-1}^{n}\prod_{j=1}^{n}\breve{\lambda}_{k|k-1}\left(x_{j}\right)dx_{1:n}\notag\\
	& =e^{-\overline{\lambda}_{k|k-1}}\sum_{n=0}^{\infty}\frac{1}{n!}\sum_{\mathbf{z}_{1}\uplus...\uplus\mathbf{z}_{n}=\mathbf{z}}\prod_{j=1}^{n}\tau_{\mathbf{z}_{j}},
\end{align}
where $\tau_{\mathbf{w}}$ is defined in (\ref{eq:tau_w}). We write $\mathbf{z}=\left\{ z^{1},...,z^{m}\right\} $, $m>1$ then
\begin{align}
	l_{k|k-1}^{T}&\left(\left\{ z^{1},...,z^{m}\right\} \right)\notag\\  =&e^{-\overline{\lambda}_{k|k-1}}\sum_{n=0}^{\infty}\frac{1}{n!}\sum_{\mathbf{z}_{1}\uplus...\uplus\mathbf{z}_{n}=\left\{ z^{1},...,z^{m}\right\} }\prod_{j=1}^{n}\tau_{\mathbf{z}_{j}}\notag\\
	=&e^{-\overline{\lambda}_{k|k-1}}\sum_{n=0}^{\infty}\sum_{d=0}^{n}\frac{d!}{n!}\left(\begin{array}{c}
		n\\
		d
	\end{array}\right)\tau_{\emptyset}^{n-d}\notag\\
	&\times\sum_{\mathbf{z}_{1}\uplus...\uplus\mathbf{z}_{d}=\left\{ z^{1},...,z^{m}\right\} :|\mathbf{z}_{_{i}}|\geq1}\prod_{j=1}^{d}\tau_{\mathbf{z}_{j}}\notag\\
	%\end{align}
	%\begin{align}
	=&e^{-\overline{\lambda}_{k|k-1}}\sum_{n=0}^{\infty}\sum_{d=0}^{n}\frac{1}{\left(n-d\right)!}\tau_{\emptyset}^{n-d}\sum_{\mathcal{Q}\angle\mathbf{z}_{k}:|Q|=d}\prod_{\mathbf{w}\in\mathcal{Q}}\tau_{\mathbf{w}}\notag\\
	=&e^{-\overline{\lambda}_{k|k-1}}\sum_{d=0}^{\infty}\sum_{\mathcal{Q}\angle\mathbf{z}_{k}:|Q|=d}\prod_{\mathbf{w}\in\mathcal{Q}}\tau_{\mathbf{w}}\sum_{j=0}^{\infty}\frac{1}{j!}\tau_{\emptyset}^{j}\notag\\
	=&e^{-\overline{\lambda}_{k|k-1}}e^{\tau_{\emptyset}}\sum_{d=0}^{m}\sum_{\mathcal{Q}\angle\mathbf{z}_{k}:|Q|=d}\prod_{\mathbf{w}\in\mathcal{Q}}\tau_{\mathbf{w}}\notag\\
	=&e^{-\overline{\lambda}_{k|k-1}}e^{\tau_{\emptyset}}\sum_{\mathcal{Q}\angle\mathbf{z}_{k}}\prod_{\mathbf{w}\in\mathcal{Q}}\tau_{\mathbf{w}},
\end{align}
where $d$ sums over the number of detected targets, and $j=n-d$. For $m=0$, we have
\begin{align}
	l_{k|k-1}^{T}\left(\emptyset\right) & =e^{-\overline{\lambda}_{k|k-1}}\sum_{n=0}^{\infty}\frac{1}{n!}\tau_{\emptyset}^{n}\\
	& =e^{-\overline{\lambda}_{k|k-1}}e^{\tau_{\emptyset}}.\notag
\end{align}

\subsubsection{Adding clutter}

The density of the measurement is the union of independent
target-generated measurements and clutter measurements. We can use the convolution formula\cite{Mahler2014RFSbook} to provide
\begin{align}
	l_{k|k-1}\left(\mathbf{z}_{k}\right) = &e^{-\overline{\lambda}^{C}}\sum_{\mathbf{z}^{T}\uplus\mathbf{z}^{C}=\mathbf{z}_{k}}l_{k|k-1}^{T}\left(\mathbf{z}^{T}\right)\prod_{z\in\mathbf{z}^{C}}\lambda^{C}\left(z\right)\nonumber \\
	=&e^{-\overline{\lambda}^{C}}e^{-\overline{\lambda}_{k|k-1}}e^{\tau_{\emptyset}}\sum_{\mathbf{z}^{T}\uplus\mathbf{z}^{C}=\mathbf{z}_{k}}\sum_{\mathcal{Q}\angle\mathbf{z}^{T}}\prod_{\mathbf{w}\in\mathcal{Q}}\tau_{\mathbf{w}}\notag\\
	&\times\prod_{z\in\mathbf{z}^{C}}\lambda^{C}\left(z\right)\nonumber\\
	%\end{align}
	%\begin{align}
	=&e^{-\overline{\lambda}^{C}}e^{-\overline{\lambda}_{k|k-1}}e^{\tau_{\emptyset}}\sum_{\mathbf{z}^{T}\subseteq\mathbf{z}_{k}}\sum_{\mathcal{Q}\angle\mathbf{z}^{T}}\prod_{\mathbf{w}\in\mathcal{Q}}\tau_{\mathbf{w}}\notag\\
	&\times\prod_{z\in\mathbf{z}_{k}\setminus\mathbf{z}^{T}}\lambda^{C}\left(z\right)\nonumber \\
	=&e^{-\overline{\lambda}^{C}}e^{-\overline{\lambda}_{k|k-1}}e^{\tau_{\emptyset}}\sum_{\mathcal{Q}\angle\mathbf{z}_{k}}\prod_{\mathbf{w}\in\mathcal{Q}}\left(\kappa_{\mathbf{w}}+\tau_{\mathbf{w}}\right),\label{eq:density_measurement}
\end{align}
where the notation $\overline{\lambda}^{C}$ is the expected number of clutter measurements. We have also applied the Lemma \ref{lem:Equality_functionals1} in the last step.

\subsection{KLD minimization}
The posterior is obtained as the application of Bayes' rule \eqref{Bayes} when the predicted density is the PPP \eqref{eq:PPP_prior}
\begin{align}\label{KLD_density}
	f_{k|k}\left(\mathbf{X}\right) & =\frac{f\left(\mathbf{z}_{k}|\mathbf{x}\right)e^{-\overline{\lambda}_{k|k-1}}\overline{\lambda}_{k|k-1}^{n}\prod_{X\in\mathbf{X}}\breve{\lambda}_{k|k-1}\left(X\right)}{l_{k|k-1}\left(\mathbf{z}_{k}\right)}.
\end{align}
The best Poisson multi-trajectory density that minimizes the KLD has the same PHD as \eqref{KLD_density} \cite{Angel2019TPHD}. This PHD can be calculated as \cite{Mahler2014RFSbook} 
\begin{align}\label{KLD_multi}
	\lambda_{k|k}\left(X\right) & =\int f_{k|k}\left(\left\{ X\right\} \cup\mathbf{X}\right)\delta\mathbf{X}\\
	& =\sum_{n=0}^{\infty}\frac{1}{n!}\int f_{k|k}\left(\left\{ X,X_{1},...,X_{n}\right\} \right)dX_{1:n}.\notag
\end{align}
We proceed to compute \eqref{KLD_multi}. We first perform a decomposition of the measurement likelihood and then the PHD calculation.

\subsubsection{Decomposition of the likelihood}

Applying the convolution formula, we can write
\begin{align}
	f\left(\mathbf{z}_{k}|\left\{ x,x_{1},...,x_{n}\right\} \right)  =&\sum_{\mathbf{z}_{1}\uplus...\uplus\mathbf{z}_{n}\uplus\mathbf{w}\uplus\mathbf{z}^{C}=\mathbf{z}_{k}}f\left(\mathbf{w}|x\right)\\
	&\times e^{-\overline{\lambda}^{C}}\prod_{j=1}^{n}f\left(\mathbf{z}_{j}|x_{j}\right)\prod_{z\in\mathbf{z}^{C}}\lambda^{C}\left(z\right).\notag
\end{align}
Expanding over the cases $\mathbf{w}=\emptyset$ and $|\mathbf{w}|>0$,
we obtain
\begin{align}
	f&\left(\mathbf{z}_{k}|\left\{ x,x_{1},...,x_{n}\right\} \right) \\
	=&f\left(\emptyset|x\right)f\left(\mathbf{z}_{k}|\left\{ x_{1},...,x_{n}\right\} \right)\notag\\
	& +\sum_{\mathbf{w}\subseteq\mathbf{z}:|\mathbf{w}|>0}f\left(\mathbf{w}|x\right)f\left(\mathbf{z}_{k}\setminus\mathbf{w}|\left\{ x_{1},...,x_{n}\right\} \right),\notag
\end{align}
which can be analogously written as
\begin{align}\label{KLD_li}
	f\left(\mathbf{z}_{k}|\left\{ x\right\} \cup\mathbf{x}\right) & =f\left(\emptyset|x\right)f\left(\mathbf{z}_{k}|\mathbf{x}\right)\\
	& +\sum_{\mathbf{w}\subseteq\mathbf{z}_{k}:|\mathbf{w}|>0}f\left(\mathbf{w}|x\right)f\left(\mathbf{z}_{k}\setminus\mathbf{w}|\mathbf{x}\right)\notag
\end{align}

\subsubsection{PHD calculation}

The PHD of the posterior is given by
\begin{align}
	\lambda_{k|k}\left(X\right) =&\int f_{k|k}\left(X\cup\mathbf{X}\right)\delta\mathbf{X}\nonumber \\
	=&\frac{\lambda_{k|k-1}\left(X\right)}{l_{k|k-1}\left(\mathbf{z}_{k}\right)}\int f\left(\mathbf{z}_{k}|\left\{ x\right\} \cup\mathbf{x}\right)e^{-\overline{\lambda}_{k|k-1}}\nonumber\\
	&\times\prod_{x\in\mathbf{x}}\lambda_{k|k-1}(x)\delta\mathbf{x}. 
\end{align}
Substituting \eqref{KLD_li} into the above equation, we can write
\begin{align}
	\lambda_{k|k}\left(X\right) =&f\left(\emptyset|x\right)\lambda_{k|k-1}\left(X\right)+\frac{\lambda_{k|k-1}\left(X\right)}{l_{k|k-1}\left(\mathbf{z}_{k}\right)}\notag\\
	&\times\sum_{\mathbf{w}\subseteq\mathbf{z}_{k}:|\mathbf{w}|>0}f\left(\mathbf{w}|x\right)\int f\left(\mathbf{z}_{k}\setminus\mathbf{w}|\mathbf{x}\right)\notag\\
	&\times e^{-\overline{\lambda}_{k|k-1}}\lambda_{k|k-1}^{\mathbf{x}}\delta\mathbf{x}\nonumber \\
	%\end{align}
	%\begin{align}  
	=&f\left(\emptyset|x\right)\lambda_{k|k-1}\left(X\right)+\frac{\lambda_{k|k-1}\left(X\right)}{l_{k|k-1}\left(\mathbf{z}_{k}\right)}\label{eq:PHD_calculation1}\\
	&\times\sum_{\mathbf{w}\subseteq\mathbf{z}_{k}:|\mathbf{w}|>0}l_{k|k-1}\left(\mathbf{z}_{k}\setminus\mathbf{w}\right)f\left(\mathbf{w}|x\right).\notag
\end{align}
Let us simplify the summation in the above expression. Using (\ref{eq:density_measurement}),
we obtain
\begin{align}
	& \sum_{\mathbf{w}\subseteq\mathbf{z}_{k}:|\mathbf{w}|>0}l_{k|k-1}\left(\mathbf{z}_{k}\setminus\mathbf{w}\right)f\left(\mathbf{w}|x\right)\nonumber \\
	& =e^{-\overline{\lambda}^{C}}e^{-\overline{\lambda}_{k|k-1}}e^{\tau_{\emptyset}}\sum_{\mathbf{w}\subseteq\mathbf{z}_{k}:|\mathbf{w}|>0}f\left(\mathbf{w}|x\right)\nonumber\\
	&~~\times\sum_{\mathcal{Q}\angle\mathbf{z}_{k}\setminus\mathbf{w}}\prod_{\mathbf{v}\in\mathcal{Q}}\left(\kappa_{\mathbf{v}}+\tau_{\mathbf{v}}\right)\nonumber \\
	& =e^{-\overline{\lambda}^{C}-\overline{\lambda}_{k|k-1}+\tau_{\emptyset}}\!\!\!\sum_{\mathcal{P}\angle\mathbf{z}_{k}}\prod_{\mathbf{v}\in\mathcal{\mathcal{P}}}\left(\kappa_{\mathbf{v}}+\tau_{\mathbf{v}}\right)\sum_{\mathbf{v}\in\mathcal{P}}\frac{f\left(\mathbf{v}|x\right)}{\kappa_{\mathbf{v}}+\tau_{\mathbf{v}}},\label{eq:PHD_calculation_auxiliary_1}
\end{align}
where in the last equality we have applied Lemma \ref{lem:Equality_functionals2}. By Substituting (\ref{eq:density_measurement}) and (\ref{eq:PHD_calculation_auxiliary_1})
into (\ref{eq:PHD_calculation1}), we can obtain
\begin{align}
	\lambda_{k|k}\left(X\right)  =&f\left(\emptyset|x\right)\lambda_{k|k-1}\left(X\right)\notag\\
	& +\frac{1}{\sum_{\mathcal{Q}\angle\mathbf{z}_{k}}\prod_{\mathbf{w}\in\mathcal{Q}}\left(\kappa_{\mathbf{w}}+\tau_{\mathbf{w}}\right)}\sum_{\mathcal{P}\angle\mathbf{z}_{k}}\prod_{\mathbf{v}\in\mathcal{\mathcal{P}}}\left(\kappa_{\mathbf{v}}+\tau_{\mathbf{v}}\right)\nonumber\\
	&\times\sum_{\mathbf{v}\in\mathcal{P}}\frac{f\left(\mathbf{v}|x\right)\lambda_{k|k-1}\left(X\right)}{\kappa_{\mathbf{v}}+\tau_{\mathbf{v}}},
\end{align}
which is equivalent to (\ref{eq:TPHD_update}), completing the proof
of Proposition 1 on the general TPHD filter update. 
\section{\label{sec:AppendixB}}
This appendix shows how to recover the update step of the standard extended target model \cite{Grans2012exPHD1,ExTPHD} and standard point target model \cite{Vo2006PHD,Angel2020TM} from the general TPHD filter update in Proposition 1.
\subsection{Standard extended target model}
In the standard extended target model, the target-generated measurement density is \cite{GransXradar2015,ExTPHD}
\begin{align}
	f\left(\mathbf{z}|x^i\right) & =\begin{cases}
		1-p^{D}\left(x^i\right)+p^{D}\left(x^i\right)e^{-\gamma\left(x^i\right)} & \mathbf{z}=\emptyset\\
		p^{D}\left(x^i\right)\gamma^{\left|\mathbf{z}\right|}\left(x^i\right)e^{-\gamma\left(x^i\right)}\prod_{z\in\mathbf{z}}l(z|x^i) & \left|\mathbf{z}\right|>0.
	\end{cases}\label{eq:standard_extended_measurement}
\end{align}
The notation $l(z|x)$ is the likelihood function for a single 
target generated measurement. By substituting \eqref{eq:standard_extended_measurement} into \eqref{likelihood}, the pseudolikelhood function becomes
\begin{align}
	L_{\mathbf{z}_{k}}&\left(x^i\right)  =1-p^{D}\left(x^i\right)+p^{D}\left(x^i\right)e^{-\gamma\left(x^i\right)}\\
	& +\sum_{\mathcal{P}\angle\mathbf{z}_{k}}w_{\mathcal{P}}\cdot p^{D}\left(x^i\right)\notag\\
	&\times\sum_{\mathbf{w}\in\mathcal{P}}\frac{\gamma^{\left|\mathbf{w}\right|}\left(x^i\right)e^{-\gamma\left(x^i\right)}\prod_{z\in\mathbf{w}}l(z|x^i)}{\kappa_{\mathbf{w}}+\tau_{\mathbf{w}}},\notag
\end{align}
where
\begin{align}
	\tau_{\mathbf{w}} =&\int p^{D}\left(x^i\right)\gamma^{\left|\mathbf{w}\right|}\left(x^i\right)e^{-\gamma\left(x^i\right)}\left[\prod_{z\in\mathbf{w}}l(z|x^i)\right]\\
	&\times\lambda_{k|k-1}\left(x^i\right)dx^i,\notag\\
	\kappa_{\mathbf{w}} =&\delta_{1}\left[|\mathbf{w}|\right]\left[\prod_{z\in\mathbf{w}}\lambda^{C}\left(z\right)\right],\quad|\mathbf{w}|>0,\\
	w_{\mathcal{P}} =&\frac{\prod_{\mathbf{w}\in\mathcal{P}}\left(\kappa_{\mathbf{w}}+\tau_{\mathbf{w}}\right)}{\sum_{\mathcal{Q}\angle\mathbf{z}_{k}}\prod_{\mathbf{w}\in\mathcal{Q}}\left(\kappa_{\mathbf{w}}+\tau_{\mathbf{w}}\right)}.
\end{align}
For simplicity, we define, for $|\mathbf{w}|>0$,
\begin{align}
	d_{\mathbf{w}}&  =\frac{\kappa_{\mathbf{w}}+\tau_{\mathbf{w}}}{\prod_{z\in\mathbf{w}}\lambda^{C}\left(z\right)}=\delta_{1}\left[|\mathbf{w}|\right]\\
	+&\int p^{D}\left(x\right)\gamma^{\left|\mathbf{w}\right|}\left(x\right)e^{-\gamma\left(x\right)}\left[\prod_{z\in\mathbf{w}}\frac{l(z|x)}{\lambda^{C}\left(z\right)}\right]\lambda_{k|k-1}\left(x\right)dx.\notag
\end{align}
Therefore, the weight $w_{\mathcal{P}}$ for each partition can be written as
\begin{align}
	w_{\mathcal{P}} & =\frac{\prod_{\mathbf{w}\in\mathcal{P}}\left(\kappa_{\mathbf{w}}+\tau_{\mathbf{w}}\right)}{\sum_{\mathcal{Q}\angle\mathbf{z}_{k}}\prod_{\mathbf{w}\in\mathcal{Q}}\left(\kappa_{\mathbf{w}}+\tau_{\mathbf{w}}\right)}\\
	& =\frac{\left[\prod_{z\in\mathbf{z}_{k}}\lambda^{C}\left(z\right)\right]\prod_{\mathbf{w}\in\mathcal{P}}d_{\mathbf{w}}}{\left[\prod_{z\in\mathbf{z}_{k}}\lambda^{C}\left(z\right)\right]\sum_{\mathcal{Q}\angle\mathbf{z}_{k}}\prod_{\mathbf{w}\in\mathcal{Q}}d_{\mathbf{w}}}\notag\\
	& =\frac{\prod_{\mathbf{w}\in\mathcal{P}}d_{\mathbf{w}}}{\sum_{\mathcal{Q}\angle\mathbf{z}_{k}}\prod_{\mathbf{w}\in\mathcal{Q}}d_{\mathbf{w}}}.\notag
\end{align}
Similarly, the pseudolikelihood function can be obtained as
\begin{align}
	L_{\mathbf{z}_{k}}&\left(x^i\right) 
	=1-p^{D}\left(x^i\right)+p^{D}\left(x^i\right)e^{-\gamma\left(x^i\right)}\\
	& +\sum_{\mathcal{P}\angle\mathbf{z}_{k}}w_{\mathcal{P}}\sum_{\mathbf{w}\in\mathcal{P}}\frac{p^{D}\left(x^i\right)\gamma^{\left|\mathbf{w}\right|}\left(x^i\right)e^{-\gamma\left(x^i\right)}\prod_{z\in\mathbf{w}}\frac{l(z|x^i)}{\lambda^{C}\left(z\right)}}{d_{\mathbf{w}}}\notag
\end{align}
which coincides with the result in \cite{ExTPHD, GransXradar2015}, as required.

\subsection{Standard point target model}

In the standard point target model, the target-generated measurement density is
\begin{align}
	f\left(\mathbf{z}|x^i\right) & =\begin{cases}
		1-p^{D}\left(x^i\right) & \mathbf{z}=\emptyset\\
		p^{D}\left(x^i\right)l(z|x^i) & \mathbf{z}=\left\{ z\right\} \\
		0 & \left|\mathbf{z}\right|>1.
	\end{cases}\label{eq:standard_point_measurement}
\end{align}
Then, equation (\ref{eq:tau_w}) becomes
\begin{align}
	\tau_{\mathbf{w}} & =\begin{cases}
		\int\left(1-p^{D}\left(x^i\right)\right)\lambda_{k|k-1}\left(x^i\right)dx^i & \mathbf{w}=\emptyset\\
		\int p^{D}\left(x^i\right)l(z|x^i)\lambda_{k|k-1}\left(x^i\right)dx^i & \mathbf{w}=\left\{ z\right\} \\
		0 & \left|\mathbf{w}\right|>1.
	\end{cases}
\end{align}
Since the point target can generate at maximum one measurement, we have that
\begin{align}
	w_{\mathcal{P}} & =0\,\exists\mathbf{w}\in\mathcal{P}:|\mathbf{w}|>1.
\end{align}
So for $\mathbf{z}_{k}=\left\{ z_{k}^{1},...,z_{k}^{m_{k}}\right\} $,
only the partition $\mathcal{P}=\left\{ \left\{ z_{k}^{1}\right\} ,...,\left\{ z_{k}^{m_{k}}\right\} \right\} $
has non-zero weights. This implies that we can write 
\begin{align}
	\sum_{\mathcal{P}\angle\mathbf{z}_{k}}w_{\mathcal{P}}&\sum_{\mathbf{w}\in\mathcal{P}}\frac{f\left(\mathbf{w}|x^i\right)}{\kappa_{\mathbf{w}}+\tau_{\mathbf{w}}} \\
	& =w_{\left\{ \left\{ z_{k}^{1}\right\} ,...,\left\{ z_{k}^{m_{k}}\right\} \right\} }\sum_{\mathbf{w}\in\left\{ \left\{ z_{k}^{1}\right\} ,...,\left\{ z_{k}^{m_{k}}\right\} \right\} }\frac{f\left(\mathbf{w}|x^i\right)}{\kappa_{\mathbf{w}}+\tau_{\mathbf{w}}}.\notag
\end{align}
In addition, we have that
\begin{align}
	&w_{\left\{ \left\{ z_{k}^{1}\right\} ,...,\left\{ z_{k}^{m_{k}}\right\} \right\} }\\ & =\frac{\prod_{z\in\mathbf{z}_{k}}\left(\lambda^{C}\left(z\right)+\int p^{D}\left(x\right)l(z|x^i)\lambda_{k|k-1}\left(x^i\right)dx^i\right)}{\prod_{z\in\mathbf{z}_{k}}\left(\lambda^{C}\left(z\right)+\int p^{D}\left(x\right)l(z|x^i)\lambda_{k|k-1}\left(x^i\right)dx^i\right)}=1.\notag
\end{align}
By substituting the previous equation and Eq. \eqref{eq:standard_point_measurement} into the general pseuodlikelihood \eqref{likelihood}, we obtain the pseudolikelihood for the standard point target TPHD filter:
\begin{align}
	\sum_{\mathcal{P}\angle\mathbf{z}_{k}}w_{\mathcal{P}}&\sum_{\mathbf{w}\in\mathcal{P}}\frac{f\left(\mathbf{w}|x^i\right)}{\kappa_{\mathbf{w}}+\tau_{\mathbf{w}}}
	=\sum_{\mathbf{w}\in\left\{ \left\{ z_{k}^{1}\right\} ,...,\left\{ z_{k}^{m_{k}}\right\} \right\} }\frac{f\left(\mathbf{w}|x^i\right)}{\kappa_{\mathbf{w}}+\tau_{\mathbf{w}}}\notag\\
	& =\sum_{z\in\mathbf{z}_{k}}\frac{p^{D}\left(x^i\right)l(z|x^i)}{\lambda^{C}\left(z\right)+\int p^{D}\left(x^i\right)l(z|x^i)\lambda_{k|k-1}\left(x\right)dx^i},\notag
\end{align}
which provides the same result for point target tracking in the TPHD filter \cite{Angel2019TPHD}.

\section{\label{sec:AppendixC}}

In this appendix, we prove Lemma \ref{lem:Equality_functionals1}.
We know that \cite[Eq. (3.6)]{Mahler2014RFSbook}
\begin{align}
	\left(\kappa+\tau\right)^{\mathcal{Q}} & =\sum_{\mathcal{A}\subseteq\mathcal{Q}}\kappa^{\mathcal{A}}\tau^{\mathcal{Q}\setminus\mathcal{A}}.\label{eq:Binomial}
\end{align}

Plugging (\ref{eq:Binomial}) into the right hand side of (\ref{eq:equality_functionals1}),
we obtain

\begin{align}
	\sum_{Q\angle\mathbf{z}}\left(\kappa+\tau\right)^{\mathcal{Q}} & =\sum_{Q\angle\mathbf{z}}\sum_{\mathcal{A}\subseteq Q}\kappa^{\mathcal{A}}\tau^{Q\setminus\mathcal{A}}.\label{eq:Binomial_sum}
\end{align}

We note that $\kappa^{\mathcal{A}}$ is different from zero only if
$\mathcal{A}$ contains single element sets $\mathcal{A}=\left\{ \left\{ z\right\} :z\in\mathbf{a},\mathbf{a}\subseteq\mathbf{z}\right\} $,
and in this case, it takes value $\kappa^{\mathcal{A}}=\lambda^{\mathbf{a}}$.
Therefore, we can first sum over all $\mathbf{a}\subseteq\mathbf{z}$
in (\ref{eq:Binomial_sum}), and then select the partitions $\mathcal{A}$
that meet this constraint. That is
\begin{align*}
	\sum_{Q\angle\mathbf{z}}\sum_{\mathcal{A}\subseteq Q}\kappa^{\mathcal{A}}\tau^{Q\setminus\mathcal{A}} & =\sum_{\mathbf{a}\subseteq\mathbf{z}}\lambda^{\mathbf{a}}\sum_{Q\angle\mathbf{z}}\sum_{\mathcal{A}\subseteq Q:\mathcal{A}=\left\{ \left\{ z\right\} :z\in\mathbf{a}\right\} }\tau^{Q\setminus\mathcal{A}}\\
	& =\sum_{\mathbf{a}\subseteq\mathbf{z}}\lambda^{\mathbf{a}}\sum_{Q\angle\mathbf{z}:\mathcal{A}\subseteq Q:\mathcal{A}=\left\{ \left\{ z\right\} :z\in\mathbf{a}\right\} }\tau^{Q\setminus\mathcal{A}}\\
	& =\sum_{\mathbf{a}\subseteq\mathbf{z}}\lambda^{\mathbf{a}}\sum_{Q\angle\mathbf{z}:\left\{ \left\{ z\right\} :z\in\mathbf{a}\right\} \subseteq Q}\tau^{Q\setminus\left\{ \left\{ z\right\} :z\in\mathbf{a}\right\} }.
\end{align*}

We make a change of variables and sum over $\mathbf{y}=\mathbf{z}\setminus\mathbf{a}$
instead of $\mathbf{a}=\mathbf{z}\setminus\mathbf{y}$. This yields
\begin{align}
	\sum_{Q\angle\mathbf{z}}\left(\kappa+\tau\right)^{\mathcal{Q}} & =\sum_{\mathbf{\mathbf{y}}\subseteq\mathbf{z}}\lambda^{\mathbf{z}\setminus\mathbf{y}}\sum_{Q\angle\mathbf{z}:\left\{ \left\{ z\right\} :z\in\mathbf{z}\setminus\mathbf{y}\right\} \subseteq Q}\tau^{Q\setminus\left\{ \left\{ z\right\} :z\in\mathbf{z}\setminus\mathbf{y}\right\} }.\label{eq:Binomial_sum2}
\end{align}

We finally have that
\begin{align}
	Q\angle\mathbf{z}:&\left\{ \left\{ z\right\} :z\in\mathbf{z}\setminus\mathbf{y}\right\} \subseteq Q \notag\\
	&=\left\{ \left\{ z\right\} :z\in\mathbf{z}\setminus\mathbf{y}\right\} \cup Q\angle\left(\mathbf{z}\setminus\mathbf{y}\right).\label{eq:set_all_partitions_equivalence}
\end{align}
That is, the set of all partitions of $\mathbf{z}$ that contain $\left\{ \left\{ z\right\} :z\in\mathbf{z}\setminus\mathbf{y}\right\} $
is equivalent to the union of $\left\{ \left\{ z\right\} :z\in\mathbf{z}\setminus\mathbf{y}\right\} $
and the set of all partitions of $\mathbf{z}\setminus\mathbf{y}$.Then, plugging (\ref{eq:set_all_partitions_equivalence}) into (\ref{eq:Binomial_sum2}),
we obtain (\ref{eq:equality_functionals1}), which finishes the proof of Lemma 1.

\section{\label{sec:AppendixD}}

In this appendix, we prove Lemma \ref{lem:Equality_functionals2}.
As $f\left(\mathbf{v}\right)$ in (\ref{eq:equality_functionals2})
is defined for all $\mathbf{v}\subseteq\mathbf{z}_{k}:|\mathbf{v}|>0$,
we can write the right-hand side of (\ref{eq:equality_functionals2})
as
\begin{align}
	& \sum_{\mathcal{P}\angle\mathbf{z}_{k}}\sum_{\mathbf{v}\in\mathcal{P}}g^{\mathcal{P}}\frac{f\left(\mathbf{v}\right)}{g\left(\mathbf{v}\right)}\nonumber \\
	& =\sum_{\mathbf{w}\subseteq\mathbf{z}_{k}:|\mathbf{w}|>0}\sum_{\mathcal{P}\angle\mathbf{z}_{k}:\mathbf{w}\in\mathcal{P}}\sum_{\mathbf{v}\in\mathcal{P}:\mathbf{v}=\mathbf{w}}g^{\mathcal{P}}\frac{f\left(\mathbf{v}\right)}{g\left(\mathbf{v}\right)}.\label{eq:equality_functions2b}
\end{align}
That is, in (\ref{eq:equality_functions2b}), we first sum all
the possible inputs $\mathbf{w}$ of $f\left(\cdot\right)$, which
implies we then only have to consider the partitions of $\mathbf{z}_{k}$
that have $\mathbf{w}$ as an element, and we constrain the sum over
the elements of each partitions accordingly. Then, we can write (\ref{eq:equality_functions2b})
as
\begin{align*}
	& =\sum_{\mathbf{w}\subseteq\mathbf{z}_{k}:|\mathbf{w}|>0}f\left(\mathbf{w}\right)\sum_{\mathcal{P}\angle\mathbf{z}_{k}:\mathbf{w}\in\mathcal{P}}\frac{g^{\mathcal{P}}}{g\left(\mathbf{w}\right)}\\
	& =\sum_{\mathbf{w}\subseteq\mathbf{z}_{k}:|\mathbf{w}|>0}f\left(\mathbf{w}\right)\sum_{\mathcal{P}\angle\mathbf{z}_{k}:\mathbf{w}\in\mathcal{P}}g^{\mathcal{P}\setminus\left\{ \mathbf{w}\right\} }\\
	& =\sum_{\mathbf{w}\subseteq\mathbf{z}_{k}:|\mathbf{w}|>0}f\left(\mathbf{w}\right)\sum_{\mathcal{Q}\angle\mathbf{z}_{k}\setminus\mathbf{w}}g^{\mathcal{Q}}.
\end{align*}
The last equality follows that the partitions of $\mathbf{z}_{k}$
in which $\mathbf{w}$ is an element are equivalent to the union of
the set $\left\{ \mathbf{w}\right\} $ and the partitions of $\mathbf{z}_{k}\setminus\mathbf{w}$ \cite{Mahler2014RFSbook}.
This finishes the proof of Lemma 2. 

\section{\label{sec:AppendixE}}
	There are different options to estimate the set of trajectories from a TPHD of the form \eqref{De_PHD}. Here, we use a sub-optimal estimator that first consists on estimating the expected number of extended target trajectories, $N_{e,k}$, and point target trajectories, $N_{p,k}$. Then, the estimator reports the start times and mean trajectories of the PHD Gaussian extended target components with the $N_{e,k}$ highest weights for extended targets. A similar procedure is done for estimating point target trajectories. 
	\par In mathematical terms, the estimation step of the G-TPHD filter is given as 
	\begin{align*}
		N_{e, k} = \text{round}(\sum_{j=1}^{J_e^{k}}{\omega_{e,j}^k})	
	\end{align*}
	for number estimate of alive extended target trajectories and 
	\begin{align*}
		N_{p, k} = \text{round}(\sum_{j=1}^{J_p^{k}}{\omega_{p,j}^k})
	\end{align*}
	for number estimate of alive point target trajectories, where both $\omega_{e,j}^{k}$ and $\omega_{p,j}^{k}$ are given by \eqref{w_e} and \eqref{w_p}. Then, the estimator reports the estimated set of trajectories
	\begin{align*}
		\mathbf{\hat{X}}_{k}=\{(t_{p,1},\widehat {m}^k_{p,1}),...,(t_{p,N_{p, k}},\widehat {m}^k_{p,N_{p, k}}),\notag\\
		,...,(t_{e,1},\widehat m_{e,1}^{k}), (t_{e,N_{e, k}},\widehat {m}^k_{e,N_{e, k}})\}\notag.
	\end{align*} 
	where ${t}_{p,i}$ and $\widehat{m}_{p,i}^{k}$ denote the start time and mean trajectory of the Gaussian density with the $i$-th highest weight for point targets, and ${t}_{e,i}$ and $\widehat{m}_{e,i}^{k}$ represent the analogous quantities for extended targets. It should also be noted that the $L$-scan approximation \cite{Svensson2014TMTT} does not result in fragmented trajectories. All estimated trajectories are reported from its start time to the current time, without fragmentation.
\bibliographystyle{IEEEtaes}
\bibliography{UN}

\begin{IEEEbiography}[{\includegraphics[width=1in,height=1.25in]{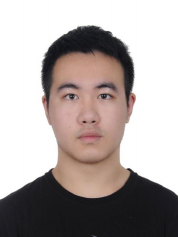}}]{Shaoxiu Wei} received the B.E. degree in communication engineering from the University of Electronic Science and Technology of China (UESTC). He is currently working towards Ph.D degree at University of California San Diego (UCSD). His research interests include statistical signal processing and multi-target tracking. \end{IEEEbiography}
\vspace{1.0cm}
\begin{IEEEbiography}[{\includegraphics[width=1in,height=1.25in]{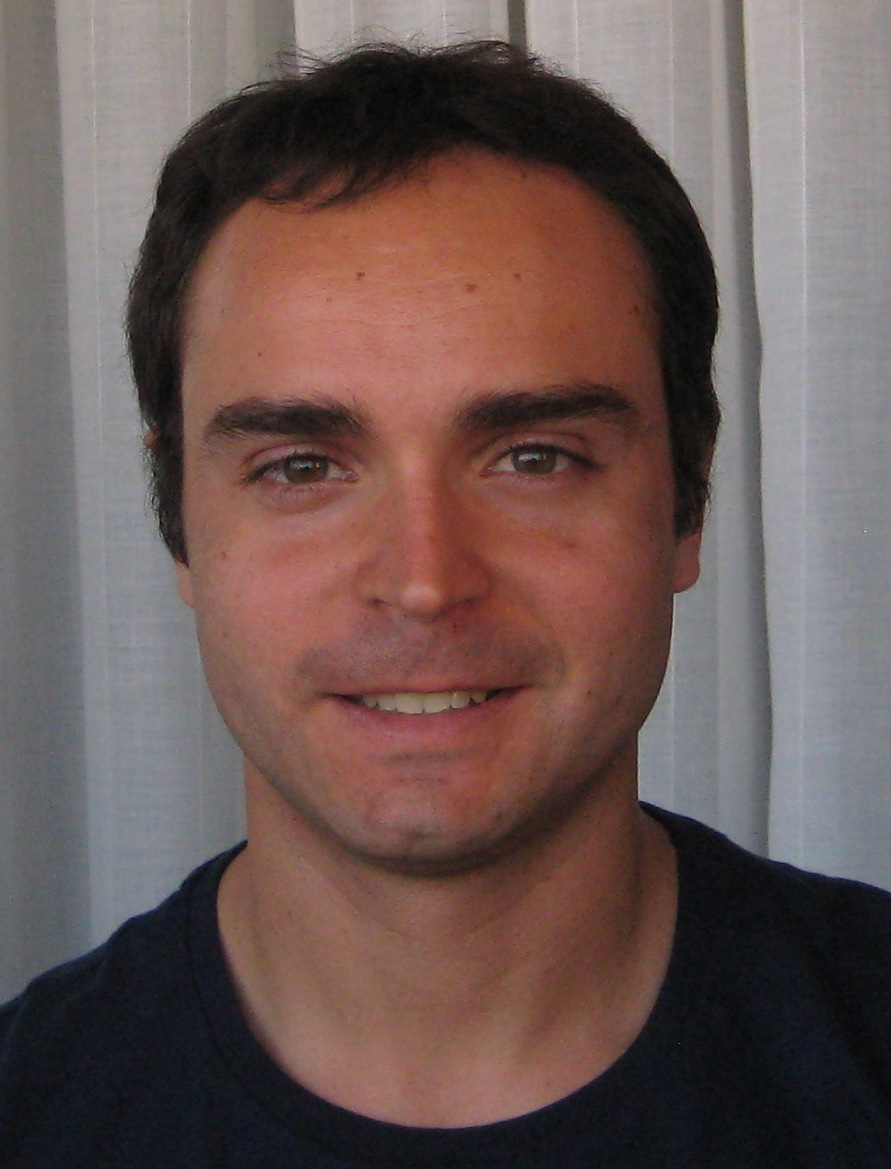}}]{\'Angel F. Garc\'ia-Fern\'andez} received the telecommunication engineering degree (with honours) and the Ph.D. degree from Universidad Polit\'ecnica de Madrid, Madrid, Spain, in 2007 and 2011, respectively. He is currently a Senior Lecturer in the Department of Electrical Engineering and Electronics at the University of Liverpool, Liverpool, UK. 

His main research activities and interests are in the area of Bayesian  estimation, with emphasis on dynamic systems and multiple target tracking. \end{IEEEbiography}
\vspace{1.0cm}
\begin{IEEEbiography}[{\includegraphics[width=1in,height=1.25in]{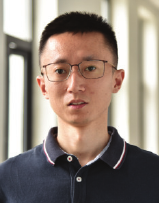}}]{Wei Yi} (M’14–SM’22) received his B.E. and Ph.D. degrees in electronic engineering from the University of Electronic Science and Technology of China (UESTC), Chengdu, China, in 2006 and 2012, respectively. From 2010 to 2012, he was a visiting student at the Melbourne Systems Laboratory, University of Melbourne, VIC, Australia.

Since 2012, he has been with the School of Information and Communication Engineering at UESTC, where he is currently a Full Professor. His research interests include target detection and tracking, radar signal processing, multi-sensor information fusion, and resource management.

Dr. Yi serves as an Associate Editor for the IEEE Transactions on Signal Processing and the IEEE Transactions on Aerospace and Electronic Systems. He is also a member of the Editorial Boards of the Journal of Radars and MDPI Sensors. He received the First Place Award in the Best Student Paper Competition at the 2012 IEEE Radar Conference in Atlanta and the Best Student Paper Award at the 15th FUSION Conference in Singapore, 2012. Additionally, he served as the General Co-Chair of ICCAIS 2019 and has been a Technical Program Committee Member for international conferences, including the IEEE Radar Conference and the FUSION Conference. \end{IEEEbiography}

\end{document}